\documentclass[journal=jacsat,manuscript=article]{achemso}

\usepackage[version=3]{mhchem} 
\usepackage{color}
\usepackage{xcolor}
\usepackage{soul}
\usepackage{wasysym}
\usepackage{stix}
\usepackage{siunitx}
\usepackage{soul}
\usepackage{epstopdf}

\author{Fernanda R. Leivas}
\affiliation{Université de Lyon, Université Claude Bernard Lyon 1, CNRS, Institut Lumière Matière, F-69622 Villeurbanne, France}
\alsoaffiliation{Universidade Federal do Rio Grande do Sul, Instituto de Física, CP 15051, 91501-970 Porto Alegre RS, Brazil}
\email{fernanda.leivas@ufrgs.br}
\author{Menghua Zhao}
\affiliation{Université de Lyon, Université Claude Bernard Lyon 1, CNRS, Institut Lumière Matière, F-69622 Villeurbanne, France} 
\author{Aymeric Allemand}
\affiliation{Université de Lyon, Université Claude Bernard Lyon 1, CNRS, Institut Lumière Matière, F-69622 Villeurbanne, France} 
\author{Cécile Cottin-Bizonne}
\affiliation{Université de Lyon, Université Claude Bernard Lyon 1, CNRS, Institut Lumière Matière, F-69622 Villeurbanne, France} 
\author{Stella {M.M.} Ramos}
\affiliation{Université de Lyon, Université Claude Bernard Lyon 1, CNRS, Institut Lumière Matière, F-69622 Villeurbanne, France}
\author{Marcia C. Barbosa}
\affiliation{Universidade Federal do Rio Grande do Sul, Instituto de Física, CP 15051, 91501-970 Porto Alegre RS, Brazil}
\author{Anne-Laure Biance}
\affiliation{Université de Lyon, Université Claude Bernard Lyon 1, CNRS, Institut Lumière Matière, F-69622 Villeurbanne, France}

\title[An \textsf{achemso} demo]
  {Condensation effect and transport on Alumina porous
membranes}

\abbreviations{IR,NMR,UV}

\begin{document}

\begin{abstract}

Understanding the adsorption of water and characterizing the water film formed within nanostructures are essential for advancements in fields such as nanofluidics, water purification, and biosensing devices. In our research, we focus on studying the condensation and transport of water through an alumina membrane with nanopores of varying wettabilities. We introduce a method to alter the membrane's wettability and enhance dissociative adsorption by varying the duration of exposure during plasma cleaning. To create different experimental environments, we modify humidity levels by controlling vapor pressure. To investigate water transport within the membrane, we apply a voltage and analyze the resulting current response. Our analysis indicates that transport properties improve with thicker water films. We  use the Polanyi theory of adsorption to capture the physics of the problem. Analyzing the conductance inside the nanopores, we find that the first monolayers may stagnate due to interactions with the pore walls. This research significantly enhances our understanding of vapor condensation within nanomaterials, particularly considering the influence of different wettabilities. These findings have broad implications for applications such as water vapor capture and related technologies.
\end{abstract}

\section{Introduction}

The characterization of water film adsorption and transport within nanopores is fundamental to understanding many physical issues such as water purification~\cite{bonilla2017adsorption,sarapat2020,Abal2021, KOHLER2018331}, energy storage~\cite{Baoxing2014, fernandez2020advances}, geological~\cite{brantley2008kinetics} and biological problems such as drug delivery or chemical sensing~\cite{wheeler2008,Cira2015, GYURCSANYI2008627, Wang2013, Bocquet2010}. The process of adsorption varies depending on factors such as temperature, humidity, and material properties like wettability, roughness~\cite{C2SM25502B, Ranathunga2020, EDALATPOUR2018967} and the hydrophobicity~\cite{ohba2021,kohler2019,zheng2019}. Condensation and transport of water at the nanoscale can then be useful for developing atmospheric water harvesting (AWH) with a low energetic cost~\cite{Leivas_JCP,Leivas_Beils}.

The vast area of nanofluidics has been attracting the attention of researchers in recent years~\cite{JIANG20223437, WEERAKOON2019335, Bocquet2010, eijkel2005nanofluidics}. Nanoconfined water is organized in layers~\cite{giacomello2021}, with the contact layer being more structured~\cite{bagchi2020} and exhibiting slower diffusional dynamics compared to those further from the surface~\cite{luca2019}. In this multilayer structure, the slip length is a fundamental parameter that controls the flow of confined water under huge pressures.~\cite{donga2017}. Although the heterogeneous dynamics of confined waters under pressure is well established, there is still a need for precise quantification of the layer thickness and its mobility at different levels of wettability.\cite{Chonghai2023, Chen2013,colloids3030055, Dette2018, PhysRevB.69.195404}. 

A convienient way to study the water mobility is the electrokinetics method that we adopt in the current study. In this case, an external electrical field is applied and the subsequent current is measured, from which we deduce the ion mobility and the water flow. Because of the strong interactions between the liquid and surface at the interface~\cite{algara2015square,giacomello2021,bagchi2020}, the nature of the confining material is expected to affect this process.  A. Allemand et al.~\cite{aymeric2023} investigated nanoscale transport properties on a silica (SiO$_2$) surface, controlling water layer thickness by adjusting relative humidity (RH). They proposed a new method to explore ionic transport under extreme molecular confinement by adjusting liquid film thickness from \SI{0.3}{\nano\meter} to \SI{2}{\nano\meter}. Their findings during conductance measurements revealed anomalous ionic transport, indicating the presence of a stagnant hydrodynamic layer in very thin liquid films, which is consistent with the low mobility of the contact layer of water under pressure. In addition to silica, alumina (Al$_2$O$_3$)  is a versatile oxide widely used as ceramic membranes due to its notable properties including high mechanical and chemical stability, good permeation performance, and electrical properties~\cite{younssi2018alumina, BENNISSAN2008223, wang2022progress, li2018highly, yang2020alumina, chen2013high,gavazzoni2017}.

In the particular case of the alumina, the charge behavior and molecular orientation in the adsorbed liquid can vary depending on the pH~\cite{ZhangJACS2008, Franks2007}. While alumina substrates exhibit a layer-by-layer adsorption process (physisorption), single crystals feature hydroxyl groups (chemisorption)~\cite{PetrikJCP2018}.  It has been reported that the $\alpha$-alumina surface, the most stable form of aluminum oxide, shows either no dissociative adsorption or a very slow process in contact with water under ambient conditions~\cite{Kirsch2014, Tong2015}. In contrast, other studies report the formation of hydroxyl groups and protonation at the interface~\cite{Elam1998, NELSON1998341,PRASETYO2019195}. According to Yujin Tong and co-workers~\cite{Yanhua2022}, $\alpha$-alumina surfaces are unreactive with respect to water adsorption, but the additional surface treatment, such as oxygen plasma cleaning, can alter surface adsorption properties~\cite{Elam1998}.

In order to circumvent the inconvenience of the presence of contaminants,  the system can be submitted to the process of plasma cleaning which involves the interaction of ions with a surface to remove undesirable molecules. In the meantime, the ionized oxygen activate some sites~\cite{Kuo-Chuan2020, YU2018383,WilliamF2019}, and renders the surface hydrophilic through oxidation~\cite{kim2015patterning, Siyuan2021}. Such a cleaning protocol on $\alpha$-alumina surfaces results in very low contact angles~\cite{WANG2022943}. Plasma cleaning is believed to facilitate dissociative adsorption at the $\alpha$-alumina/water interface. The plasma cleaning, therefore, can be used to control the wettability of the system.

In this paper, we provide a quantitative analysis  of the water mobility for different wettabilities. The   water adsorption and transport properties  confined in an alumina nanoporous membranes are studied under varying conditions of wettability and relative humidity. The contact angle was controlled by adjusting the exposure time of plasma cleaning treatment. The experiment was carried inside a vacuum chamber and relative humidity is controlled by adjusting vapor pressure. Ionic transport in water was studied by monitoring the current response when applying a sinusoidal voltage difference of amplitude $0.8$V. 

Here we introduced a novel application of plasma cleaning to control wettability and induce dissociative adsorption on a relatively non-reactive surface such as alumina porous membranes. By applying a potential difference across the membrane, we discovered that the current response of the film formed within the pores is more pronounced at higher levels of hydrophilicity and relative humidity. This suggests enhanced transport as the film thickness increases. 

Additionally, we observed that Polanyi's theory~\cite{polanyi1916adsorption} effectively explains the physics involved, revealing an increase in adsorption energy with greater hydrophilicity. The use of the electric field instead of a pressure allows us to  investigate the conductivity and conductance as function of the thickness of the water layer.

The remainder of the paper proceeds as follows: the next part presents the methods employed in this study, followed by the results and discussions. Conclusions finalize the manuscript.

\section{Materials and Methods}

\subsection{Membrane properties}

Anodisc aluminum oxide membranes of \SI{55}{\micro\meter} thickness purchased from Sigma-Aldrich were used as substrate. The substrates were pierced by  nanometric conical pores, randomly distributed, with diameters of \SI{200}{\nano\meter} and \SI{100}{\nano\meter} (on the opposite side) and a mean inter-pore distance of \SI{320}{\nano\meter}. These geometrical characteristics lead to a porous fraction, defined as the ratio between pore area and total membrane area, of 0.3 on the 200 nm side.

To perform conductivity measurements throughout the membrane, a thin platinum film of \SI{20}{\nano\meter}  thickness was deposited with an electron gun evaporator (Alliance Concept-model EVA300) on the surface with the larger \SI{200}{\nano\meter} pores without causing any obstructions inside the pores (see Figure~\ref{Figure1}b). The number of pores, $N_p$, surrounded by platinum (on the metalized zone) is determined by the nanopore density multiplied by the corresponding area $\approx$ \SI{1.8}{\cm^2}, $N_p$ =  $1.7 \times 10^9$ $\pm$ $0.1$.

Roughness measurements were conducted using an atomic force miscroscope (AFM, Asylum Research, MFP3D) operating in tapping mode. A zone of \SI{4}{\micro\meter^2} was scanned on both side of the membrane, the RMS roughness, defined as the average height deviations across a surface, was found to be $\approx$ \SI{50}{\nano\meter}  on the \SI{200}{\nano\meter} side and $\approx$ \SI{450}{\nano\meter} on the \SI{100}{\nano\meter} side. To probe the roughness inside the pore, one sample was cut cut in half, and a profile line scan (\SI{0.5}{\micro\meter}) was performed along the wall. In this region, the maximum peak-to-valley height (Rz) was found $<$\SI{0.2}{\nano\meter} indicating a relatively smooth wall.

\subsection{Conductance Experiment}
\label{conduc_exp}

\begin{figure}[ht!]
\includegraphics[width=15cm]{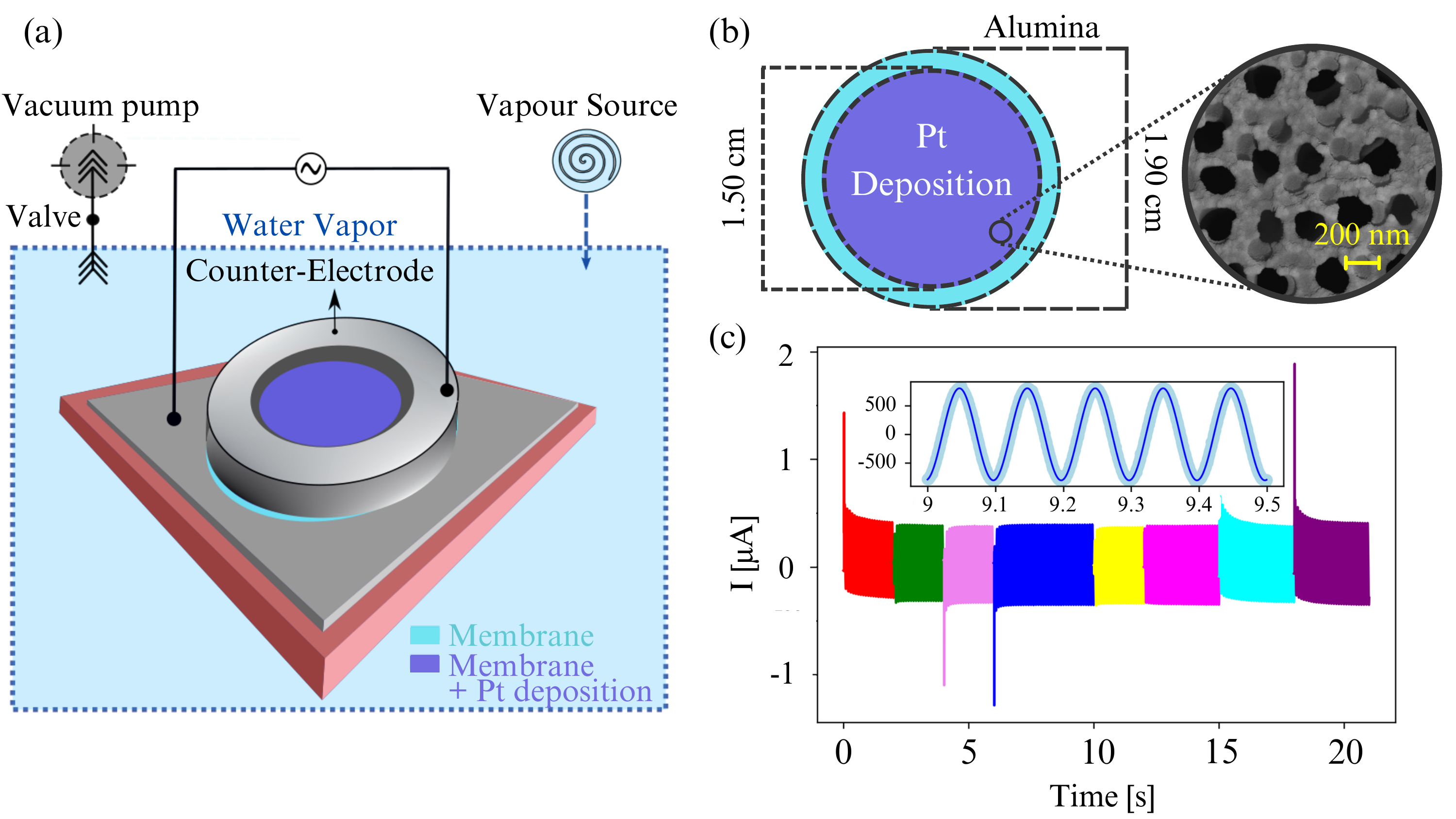}
\centering
\caption{(a) Scheme illustrating the set-up used for conductance measurement under controlled vapour pressure.
(b) Scheme and SEM picture of the Anodisc alumina membrane. Black spots correspond to nanoporous zone. The membrane diameter is \SI{1.9}{\centi\meter}, and a metallic deposition, gray zone on SEM image, was applied with a diameter of \SI{1.5}{\centi\meter}. The nanopores are randomly distributed with an average distance of \SI{320}{\nano\meter}.
(c) Electrical current measured as a function of time (RH $=90\%$, $\Theta=5^\circ\pm 1^\circ $). The input signal was transmitted and interrupted eight times, represented on the graph by different colors. Inset: Time zoom of the current response I (\SI{}{\micro\ampere}) as a function of time. The black line corresponds to a sinusoidal fit, $I(t) = A \sin({20\pi t+\theta})+c$. The resulting current is determined as the average of the resulting amplitudes ($A$) from these eight fits.}
\label{Figure1}
\end{figure}

This experiment aims to explore the transport properties of condensed water within hydrophilic nanopores, using the electrical conductivity as a key parameter. It was carried out under varying humidity levels RH = [$60\%-100\%$] and contact angle values $\Theta=[5^\circ-60^\circ]$, both of which are crucial for understanding water adsorption mechanisms~\cite{HU2019249, CihanA2019, URTEAGA2019407, C001349H, XU2020118297}. 

The experimental setup described elsewhere \cite{aymeric2023} involves a chamber (TS102V-F2, Instec) connected to a vapor source containing distilled water ($18.2$ M$\Omega$·cm, PURELAB flex, ELGA) and a vacuum pump (XDS10, Edwards). The vacuum condition is crucial for preserving the physio-chemical properties of the membrane-water interface and preventing airborne contamination. In the chamber, the membrane is positioned between an electrode and a counter electrode, as depicted in Figure~\ref{Figure1}a.

The system initially undergoes a vacuum process for the purpose of self-cleaning and is then connected to the water vapor source, which is maintained at a constant temperature of \SI{288.15}{\kelvin} using a precision thermal bath (200F, Julabo) and consequently sets the actual vapor pressure inside the chamber. The vacuum pump remains connected during the whole measurement, controlling suction flow with a metering valve (NY-2M-K6, Swagelok) with a small opening of \SI{5.5}{\micro\meter} to maintain low-pressure conditions and temperature stability. The temperature of the membrane is precisely controlled by a high-precision thermal stage (Mk2000, Instec) that is in direct contact with the membrane to ensure the high heat conductivity.

Concerning the electrical components, the side of the membrane without Pt deposition is in contact with a bottom electrode, a thin glass plate covered with 20 nm of platinum  deposited with an electron gun evaporator. The electrical connnection to the Pt covered glass is established using silver conductive paint (RS 186-3600, RS PRO) and epoxy adhesive (EA 9492, Henkel). The counter-electrode, on the membrane covered with platinum, takes the form of a metal piece with the shape of a donut, with its central portion left empty to allow water condensation on the porous membrane (see Figure~\ref{Figure1}a).

The donut has a total diameter matching that of the membrane (\SI{1.9}{\cm}), with its inside diameter corresponding to the size of the metallic deposition (\SI{1.5}{\cm}). It has a thickness of \SI{2}{\mm} and includes an additional ledge with a diameter and thickness of \SI{0.5}{\mm}, and this ledge is located at the internal radius of the donut and is designed to optimize pressure along the border of the metallic deposition, enhancing contact.

The regulation of RH is achieved by tuning the vapor pressure, with temperature acting as intensive variable following the Clausius-Clapeyron relation~\cite{pippard1964elements}: $\ln(P/P_{sat})=-\Delta H/(RT)$. Here $P$, $\Delta H$, $R$ and T are the vapor pressure, the heat of vaporization, the gas constant and the absolute temperature. By manipulating the vapor pressure through temperature, we can precisely adjust RH, given that it is expressed as $P/P_{sat}\times 100$.  
In practice, the vapor pressure inside the chamber was simultaneously monitored using a capacitance gauge (CMR362, Pfeiffer Vacuum), and saturated vapor pressure is calculated using the IAPWS formulation from 1995~\cite{Wagner2002}.

\subsubsection{Wettability Control}

Wettability properties were determined using the standard sessile drop method, which involves placing a small droplet on a substrate and analysing the droplet's shape to extract geometrical parameters of the liquid/solid interface, such as the contact angle $\Theta$. For such experiments, the samples were placed within a temperature-controlled chamber, and ultrapure water droplets of approximately \SI{2}{\micro\liter} in volume 
were deposited on the membrane's surface. The side of the membrane set to measure $\Theta$ was the one with $200$ nm pore diameter. Side-view images of the drops were recorded with a CCD camera operating with a scan rate of 15 frames per second (fps) for subsequent contact angle measurements. The values of $\Theta$ were measured in two configurations: membrane with Pt deposition and without Pt deposition. For both cases the experiment was repeated at least three times with three different membranes. The values obtained were $93^{\circ}\pm 8^{\circ}$ and $56^{\circ} \pm 4^{\circ}$, respectively.

Our focus is on water flow within nanopores in direct contact with alumina. Inside the pores the contact angle is smaller than on the Pt surface, and water will preferentially condense on the nanopore walls. From now on, when discussing wettability, we refer specifically to the contact angle of water with Al$_2$O$_3$, thus contact angles will be measured on samples without metal deposition.

To change the wettability inside the pore, the substrates have encompassed a plasma treatment (Harrick Plasma PDC-32G at high power, \SI{18}{\watt} applied to the RF coil, with compressed air as gas). Indeed, the removal of surface impurities through ionized gas reduces the contact angle~\cite{Isabell1999, banerjee2010molecular}. Figure~\ref{Figure2}a illustrates the enhancement of hydrophilicity by displaying droplet images on the membrane before and after one minute of plasma treatment.

To modify the wettability of the substrate, both the plasma exposure time and the waiting duration between treatment and sample use (relaxation time) can be varied. Studies on the temporal evolution of $\Theta$ for water on alumina after plasma treatment were carried out to adjust this parameter according to our needs. Figure~\ref{Figure2}b shows a rapid decrease in $\Theta$ from $58^\circ$ to $20^\circ$ within $10$ seconds of plasma treatment. However, this hydrophilic transformation is not permanent, as shown for example for PDMS samples, which regains their original wettability within 30 minutes when exposed to air~\cite{ZHAO201233}. 

Our samples exhibit a more extended recovery period; Figure~\ref{Figure2}c illustrates the evolution of the contact angle, $\Theta$, after a one-minute plasma treatment. It indicates that after one day, the sample recovers around $10\%$ of its initial $\Theta$ value. Complete angle recovery occurs approximately 10 days after plasma exposure. To achieve different contact angles, we then utilized the two mentioned parameters: the duration of exposure to plasma (Figure~\ref{Figure2}b), and the relaxation time after a one-minute plasma cleaner treatment (Figure~\ref{Figure2}c).

Experimentally, two membranes underwent the same plasma cleaner treatment – one with Pt deposition and one without. The membrane with metallic deposition was used for conductivity measurements, and the other served as a reference for determining the contact angle (sessile drop method) on the water-Al$_2$O$_3$ interface. Despite the slow recovery time,  to ensure that the angle measured by the reference membrane mirrors the wettability within the pores, the membrane used in the experiment was promptly introduced into the vacuum chamber after $\Theta$ measurements.

\subsubsection{Conductance measurements}
A sinusoidal biased voltage with peak to peak amplitude of $0.8$ V at \SI{10}{\hertz} was applied across the electrodes and such a selection of the voltage amplitude is to avoid electrochemical reactions. The data are sampled at a rate of \SI{2000}{\hertz}. The sinusoidal voltage, which avoids accumulation of ions, will induce a current circulation along the condensed film. This current can be consequence of the formation of hydroxyl groups at the water-Al$_2$O$_3$  interface facilitated by the activation of sites due to plasma cleaner treatment~\cite {Yanhua2022, WANG2022943, PRASETYO2019195, ZhangJACS2008, PetrikJCP2018}, among others. 

The induced current was measured using an amplifier and detected on the computer with a home-made I/V converter. Both excitation and consequential signals were remotely controlled by a Python script that operated a Digital-Analog card (NI USB-6216, National Instruments). The measured current signal is a standard sinusoid wave that shows a phase shift to the excitation stimulation and it was further fitted by a sinusoidal function. The amplitude obtained from the fit provides us the current response value from the water film condensed on cone walls. An example of the signals fitted is shown in Figure~\ref{Figure1}c. 

To ensure the reproducibility of the detected signal, the sinusoidal voltage was cycled on and off eight times along $25$s (represented by different colors in Figure~\ref{Figure1}c), and the amplitude of the induced current was calculated as the mean of these responses. The dispersion of the values in reference of the mean is calculated by standard deviation, using pandas library in python. At fixed wettability, conductance of the adsorbate liquid was systematically detected by continuously varying RH from $60\%$ to $100\%$. The recovery time between each measurement was at least $2$ minutes.

Despite a strict control on the vacuum condition, the system leakage was shown to be unavoidable~\cite{aymeric2023}. To quantify this contribution, we performed the experiment under very low humidity conditions (RH=$15\%$), where the observed current response is taken to be leakage conductance since the possible ions present in the system are immobile under this humidity~\cite{sarfati2020temporally,aymeric2023}. The gray band in Figure~\ref{Figure2}a delineates the region corresponding to the system's leakage response (the noise), where the measurements are not significant.

\subsection{Water Adsorption Experiment}

To track the effect of confinement on water transport, one needs to estimate the amount of water condensed on the pore pore walls inside the membrane. We conducted an experiment aiming at investigating the quantity of water adsorbed by the membrane in a humidity-controlled environment. To do that, we monitored the weight of the membrane thanks to a high-resolution balance (MSE225P-Sartorius) with a precision of $0.01$mg placed at various RH levels~\cite{Thomas202339}. The device of the balance includes a cubic glass chamber designed to ensure the high precision of the measurements. To adjust the relative humidity, we placed vessels with water inside the chamber. Once the chamber is sealed, water evaporates, thereby increasing continuously the relative humidity. Quantities including RH, temperature and weight were monitored and controlled using a combined hygrometer / thermometer (VOLTCRAFT HT-200) with a precision of $0.1\%$  and 1 degree. Initially, RH was at ambient value (RH$\sim 50\%$), and the initial readings were disregarded to allow the system to stabilize. The actual measurements started when RH reach $60\%$, serving as the reference humidity value and zero point of weight measurements.

The procedure unfolded as follows: initially, a plastic surface was used as a tare, and we recorded its weight ($T_{RH}$) for RH continuously increasing from $60\%$ to $80\%$. Then, the surface was dried using compressed air. A second experiment was conducted later, this time with five membranes deposed on the plastic surface ($W_{RH}$). Before starting to record $W_{RH}$, we weighed the plastic surface without the membranes. We attached the membranes only if the plastic surface weight matched with the first value measured in the previous experiment. The mass of water adsorbed at RH by one membrane, $\Delta W_{RH}$, is calculated as the difference in weight between the two scenarios divided by 5, minus the mass of water adsorbed at RH $= 60\%$, which serves as the initial measurement point.
\begin{equation}
    \Delta W_{RH} = \frac{W_{RH} - T_{RH}}{5}-\Delta W_{60\%}, \hspace{2cm} RH \in [60\%,80\%]
\end{equation}
This procedure was repeated at least four times for two extreme cases of wettability; when the membranes experienced ($\Theta = 8^\circ\pm 3^\circ$) or not ($\Theta = 58^\circ\pm 5^\circ$) one minute of plasma exposure. All experiments were conducted at a temperature T $\approx$ 300K, resulting in $\Delta W_{60\%}$ of $0.28 \pm 0.02$\SI{}{\gram} for the contact angle $\Theta = 8^\circ\pm 3^\circ$ and $0.26 \pm 0.02$\SI{}{\gram} for $\Theta = 58^\circ\pm 5^\circ$. The maximum RH we could achieve, RH=$80\%$, results from the limited size of the balance chamber, which can only hold a limited number of water vessels.

\section{Results and discussion}

\begin{figure}[ht]
\includegraphics[angle = 0, width=14cm]{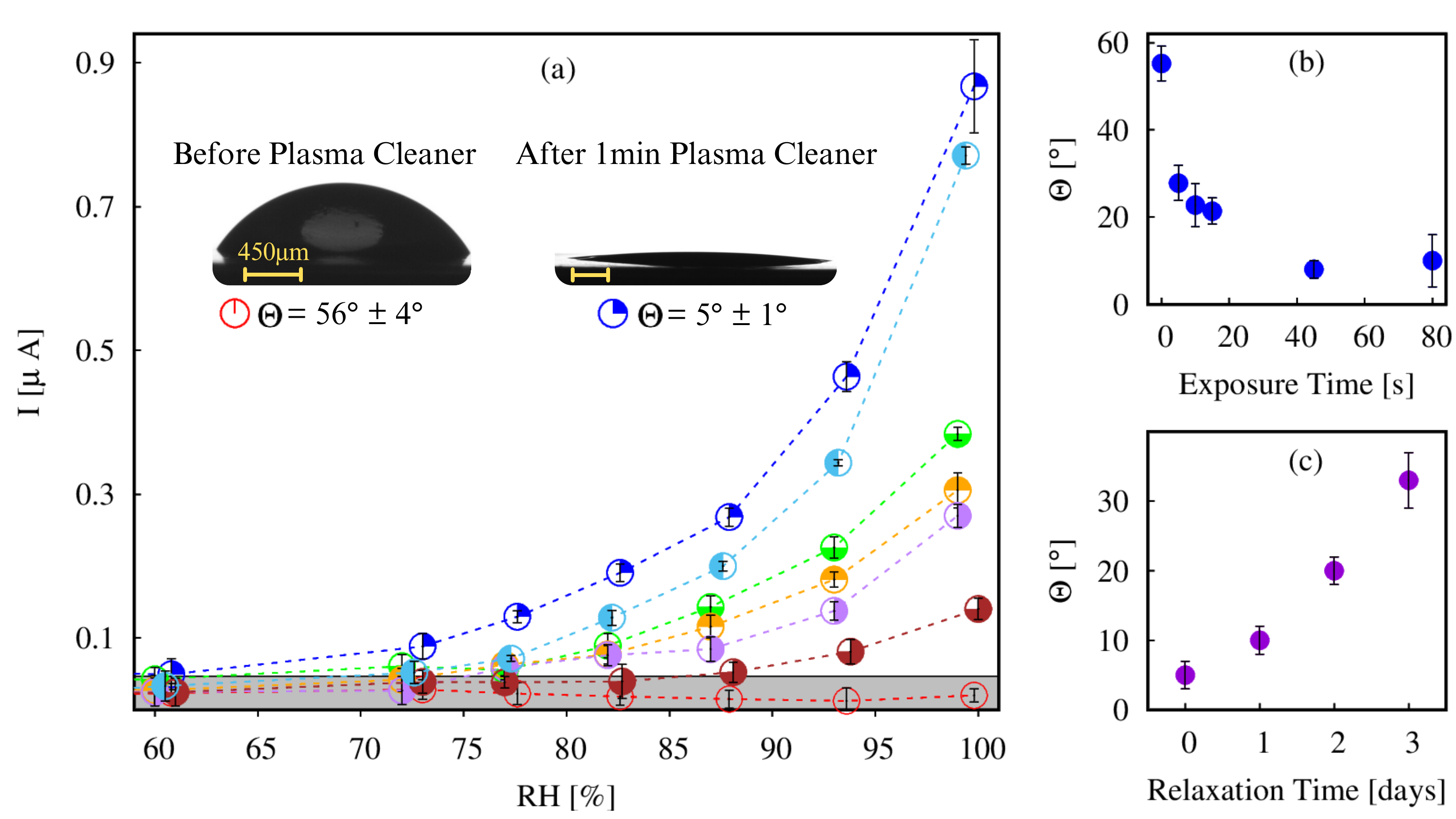}
\centering
\caption{(a) Measured current (I [\SI{}{\micro\ampere}]) as a function of RH for various conditions. The gray band at the bottom of the graph outlines the noise region. The contact angles of alumina with water, measured for each wettability were \Circle: $\Theta= 56^\circ \pm 4^\circ$, $\blackcircleulquadwhite$: $\Theta= 28^\circ \pm 3^\circ$, $\RIGHTcircle$: $\Theta= 19^\circ \pm 4^\circ$, $\circletophalfblack$: $\Theta= 13^\circ \pm 2^\circ$, $\circlebottomhalfblack$: $\Theta= 11^\circ \pm 1^\circ$, \LEFTcircle: $\Theta= 9^\circ \pm 1^\circ$, and $\circleurquadblack$: $\Theta= 5^\circ \pm 1^\circ$. Inset : Instantaneous image (side-view) of a typical drop before and after exposure to plasma cleaner. (b) Measured contact angles ($\Theta$ degres) of a water droplet on the alumina substrate as a function of the exposure time under plasma. (c) Measured contact angles ($\Theta$) of a water droplet deposited on the alumina substrate as a function of the relaxation time after exposure to one minute of plasma.}
\label{Figure2}
\end{figure}

Electrokinetic measurements were performed for seven different wettabilities with RH values ranging from $50\%$ to $100\%$, they are summarized in Figure~\ref{Figure2}a. Figure~\ref{Figure2}b and c depict the contact angles obtained after different periods under plasma treatment and by relaxation time after one minute of plasma exposure (check the details in the subsection wettability control). Analysing the curves on Figure~\ref{Figure2}a, we observe a noisy background from the membrane before plasma cleaning, suggesting that there is no dissociative adsorption. Subsequently, after plasma cleaning, as the hydrophilicity increases, so does the current response. In addition, the current also increases when the RH rises. 

The above findings can be intutively understood based on previous reports. Increasing wettability increases the thickness of the adsorbed water layer, as indicated in the research of Guo et al. for metallic materials~\cite{GUO20129087}. For water adsorption and transport on flat alumina surfaces, it has been established that the positive correlation between the electrokinetic current and the humidity is associated with the increasing thickness of adsorbed water layers~\cite{BenIEEE1987}. Regarding our results with porous alumina membranes, 
the higher induced current  with increasing  wettability and humidity may also be related to adsorption properties, such as the thickness of the water film adsorbed inside the pores. Our objective is then to determine the thickness of the adsorbed layer as a function of RH.

\subsection{Water adsorption}
In hydrophilic nanopores with diameters over $8$ nm, sorption happens from the walls towards the center (radial-pore-filling mode)~\cite{Llave2012, grunberg2004hydrogen}. The critical humidity for the transition from multilayer adsorption to capillary condensation (supersaturation) varies with pore diameter. For pores larger than $100$ nm, the effect of confinement is negligible and this transition starts from the saturated point RH = $100\%$~\cite{LI2017253}. This is indeed the case for our nanoporous membrane, and then, the pore-filling process follows a multilayer scenario, where the thickness of the water film increases with humidity~\cite{Freund1999}.

There are several models to study the adsorption of gases by surfaces. The BET (Brunauer– Emmett– Teller) model ~\cite{BET1938} is often used n the case of porous surfaces to describe multilayer adsorption, but it only works well in the range of relatively low partial pressures ($P/P_0$) typically between $0.05$ and $0.35$~\cite{NAONO1991405}, which is out of our experimental range. In our experimental range, the Polanyi theory~\cite{polanyi1916adsorption} appears more relevant, that considers the interactions between the gas molecules and the surface of the adsorbent materials via the van de Waals attraction. The application of this approach for water and porous alumina interactions has already been reported~\cite{OUCHI2014219}. The Polanyi equation, which then directly links the water adsorbed density with the RH, is expressed as~\cite{book_interf} 
\begin{equation}
    \Gamma  = \frac{1}{V_m^L}\Bigg(\frac{C}{RT\log(1/RH)}\Bigg)^{1/3} - \frac{D_0}{V_m^L},
    \label{Polanyi}
\end{equation}
where $\Gamma$ is the number of moles of water adsorbed per unit of surface, which can be expressed as the ratio between the thickness of the adsorbed layer and the molar volume of the liquid, $\Gamma = h_{RH}/V_m^L$, $D_0$ is the molecular radius of the adsorbate (gas), $V_m^L$ the molar volume of the liquid and $C=\pi\rho_{A}C_{AB}/3$ is a constant derived from Gibbs free energy for the van der Waals interaction (A is the adsorbate and B is the adsorbent). R is the constant of perfect gases and T=\SI{297}{\kelvin} the temperature. For water $V_m^L = 18\times 10^{-6}$\SI{}{\meter^3/\mol}, and $D_0=$ \SI{0.135}{\nano\meter}.

We will use the isotherm equation (\ref{Polanyi}) to determine the amount of adsorbed water, as a function of humidity in our experiments (see experimental section for details). We have to stress here that our measurements of water adsorption as a function of humidity are not absolute, but relative to the value at RH=60$\%$, and we then need a model to get the absolute value of adsorbed water.

In details, after measuring the mass adsorbed by the membrane $\Delta W_{RH}$, we can deduce the water layer thickness $h_{RH}$ formed inside the pores, if we assume it is homogeneously distributed. The relative humidity was varied between $60\%$ and $80\%$, ensuring a multilayer adsorption regime. We remind that the experiment RH$=60\%$ is taken as a reference point, or point zero of the measures. Thus, the water layer thickness $h_{RH}$ is the sum of thickness at this reference point, RH$=60\%$, plus the thickness deduced from the balance measurements $h'= \Delta W_{RH}/2\pi r_{np} L N_{Ap}\rho_{w}$. This reads:

\begin{equation}
h_{RH} =  h_{60\%} + h' \hspace{1cm} \rm{RH} = [60\%-80\%],
    \label{thickness}
\end{equation}
with the water density $\rho_{w}=0.997\times 10^{-21}g/$nm$^3$, $r_{np}=75$ nm the nanopore average radius, $L=55$ \SI{}{\micro\meter} the membrane thickness, and $N_{Ap}\approx 2.7 \times 10^9$ the number of the pores in the entire membrane (including the section without Pt deposition).

\begin{figure}[ht]
\includegraphics[width=15cm]{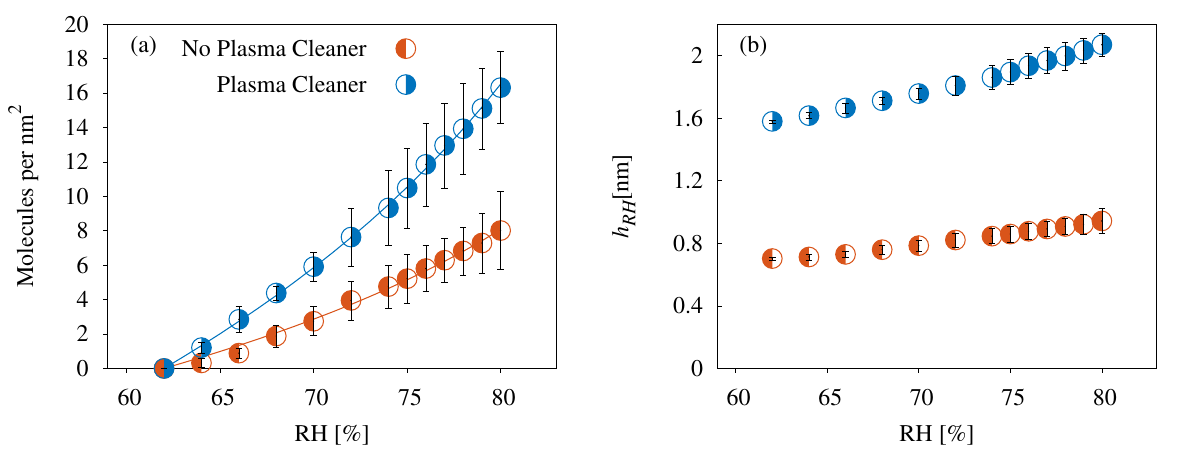}
\centering
\caption{(a) The number of molecules adsorbed per unit of surface ($\Gamma'$) taking RH$=60\%$ as reference point for RH as a funcion of humidity in the range of $[60\%-80\%]$. The data is an average of five experiments, and error bars illustrate the resulting dispersion. The line is a fit using equation \ref{Polanyi_fit}. (b) Water layer thickness ($h_{RH}$) condensed inside the nanopores using expression~\ref{thickness} as a function of humidity in the range RH$=[60\%-80\%]$.} 

\label{Figure3}
\end{figure}

Along with $h_{RH}$, the adsorption density $\Gamma$ has two components: $\Gamma_{60\%}$ at RH$=60\%$, which can be derived using equation~(\ref{Polanyi}), that sets the background, and the additional adsorption when RH$>60\%$, $\Gamma'= h'/V_m^L $. Figure~\ref{Figure3}a shows the number of molecules per area, $\Gamma'$, 
as a function of RH for two cases with and without one minute's plasma treatment; the points are deduced from experimental data, and the line represents the fit described by $\Gamma' = \Gamma - \Gamma_{60\%}$ (equation \ref{Polanyi_fit}) , with $C$ as the fitting parameter:

\begin{equation}
        \Gamma' = \frac{1}{V_m^L}\Bigg(\frac{C}{RT\log(1/RH)}\Bigg)^{1/3} - \frac{1}{V_m^L}\Bigg(\frac{C}{RT\log(1/0.6)}\Bigg)^{1/3}.
    \label{Polanyi_fit}
\end{equation}

The values of $C$ obtained from the fit are $C_n=$ \SI{690} {\joule\nano\meter^3/\mol} and $C_{pc}=$ \SI{5930} {\joule\nano\meter^3/\mol} for the case without and with plasma activation, respectively. This parameter is related to the strength of van der Waals forces between adsorbent and adsorbate, and is proportional to the Hamaker constant $A = 3\pi C\rho_A$, which quantifies the strength forces between particles or surfaces. By evaluating this Hamaker constant from $C_n$ and $C_{pc}$ we obtain $A_{n}=3.6\times 10^{-19} $\SI{}{\joule} and $A_{pc}=3.1\times 10^{-18} $\SI{}{\joule}, in fair agreement with values reported from the literature \cite{visser1972hamaker}. Moreover, the Hamaker constant for the plasma-activated case is larger than the non-activated case, ($A_{pc}> A_n$), indicating that the adsorption force between gas molecules and the surface has increased after the plasma treatment. As previously mentioned in the text, the plasma cleaning process uses ionized gas to treat the surface with plasma ions. After this process, the creation of active sites enhances surface reactivity~\cite{Alam_2014} and adsorption forces, potentially increasing the Hamaker constant. This promotes also dissociative adsorption at the alumina/water interface. As observed earlier, there is no measured ionic current when the membrane has not been treated with a plasma cleaner (Figure~\ref{Figure2}a).
Using the value found of the constant C we can then estimate the absorbed thickness $h_{60\%}$ at RH = $60\%$. For the case without plasma procedure the thickness obtained is $h_{60\%}\approx 0.7$\SI{}{\nano\meter}, and for the case with the lowest contact angle $h_{60\%}\approx 1.6$\SI{}{\nano\meter}. In Figure~\ref{Figure3}b, we show the absolute thickness $h_{RH}$, using the values found for $h_{60\%}$ and $h'$.

\subsection{Ionic transport}

\begin{figure}[ht]
\includegraphics[angle = 270, width=16cm]{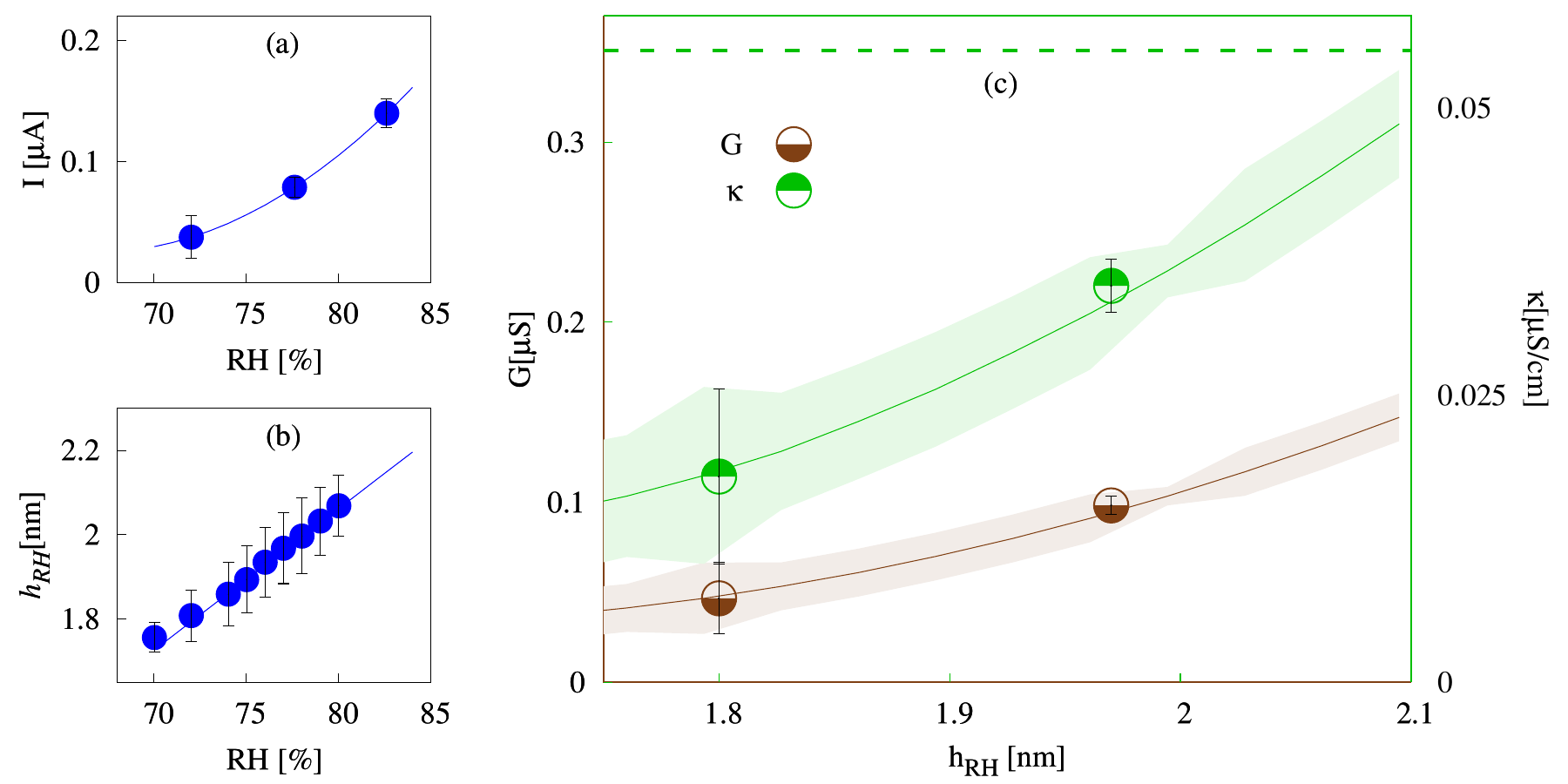}
\centering
\caption{(a) Current measured for $\Theta = 5^\circ \pm 1^\circ$ within the RH range of [70\%-85\%], corrected by subtracting the noise value (\SI{0.051}{\micro A}). The line represents an extrapolation using the quadratic equation $f(x) = ax^2+bx+c$ where $a=4.6\times 10^{-4}$ \SI{}{\micro I} per $\% ^2$, $b=-6\times 10^{-2}$ \SI{}{\micro I} per $\%$ and $c=2$ \SI{}{\micro I}.  (b) Water layer thickness within the same RH range. The line is a extrapolation using a linear function $f(x) = ax + b$ where $a=3\times 10^{-2}$ \SI{}{\nano \meter} per $\%$ and $b=-0.6$ \SI{}{\nano \meter}. (c) Left y-axis, brown: Estimated conductance as a function of the number of monolayers for the case $\Theta \approx 5^\circ$. The line was obtained using equation~\ref{conductance}, by extrapolating current I (a) and thickness $h_{RH}$ (b). Data points were obtained from the same equation directly correlating the I (a) and $h_{RH}$ (b) data points under the same humidity. Right y-axis, green: Estimated conductivity as a function of the number of monolayers also for $\Theta \approx 5^\circ $. The line was obtained by the same extrapolation (a) and (b), but using equation~\ref{kappa} for $L=$\SI{55}{\micro\meter}, $S = 2\pi  r_{np}\times h_{RH}$, $r_{np}=$\SI{75}{\nano\meter} and $N_p=1.7\times 10^9$. The points were obtained using the data points from (a) and (b) at the same RH, using equation~\ref{kappa}. The dashed line illustrates the conductivity value of the bulk distilled water $\kappa_w = 5.5 \mu$S/m. The shaded areas behind the lines indicates extrapolation errors, which include errors from existing data points and an average error estimated using the mean of all measured errors, applied to the extrapolated points.}
\label{Figure4}
\end{figure}

Moving forward, we estimate the conductivity and conductance in the adsorbed film, to investigate transport properties in our membrane. Therefore, our focus will be on the case treated with the longest plasma ($\Theta = 5^\circ \pm 1^\circ$, Figure~\ref{Figure2}a), as it shows a higher current response and reactivity. We inspect the data on the current $I$ for this wettability (Figure~\ref{Figure4}a), and we analyze the conductance trend ($G$) using the following equation:
\begin{equation}
    G = \frac{I}{V}
    \label{conductance}
\end{equation}
where $V$ is fixed at 0.8V. Since we have the values of $G$ and $h_{RH}$ for different RH, we can directly correlate them as shown on Figure~\ref{Figure4}c. The data point corresponds to experimental data using this procedure and that is obtained from the thickness and current values measured for precisely the same RH. Solid lines correspond to values obtained from extrapolations made within the RH range of $70\%$ to $85\%$, the range in which experimental measurements for both $I$ and $h_{RH}$ coexist. It consists of a quadratic fit for I(RH) measurements and a linear fit for $h_{RH}$(RH) data, as reported in Figures~\ref{Figure4}a and b. 

As the thickness increases in Figure~\ref{Figure4}c (left axis), the conductance also increases. This correlation may be a consequence of the fact that in the initial layers the interaction forces between the liquid and the wall are very strong. As the liquid fraction expands, these forces decrease with distance from the surface, thus facilitating conductance.

To better understand this result, we further study the trend of conductivity $\kappa$ by extrapolating $I$ and $h_{RH}$ as we did for $G$, using the relation
\begin{equation}
    \kappa = G\times\frac{L}{SN_p}
    \label{kappa}
\end{equation}
where cross-sectional of area of the conducting water layer is given by $ S= 2\pi r_{np} \times h_{RH}$, and $L$ is the distance between the electrodes (length of pores/membrane thickness). In Figure~\ref{Figure4}c (right axis) the resulting conductivity trend is presented, along with data points, the dashed line represents the conductivity of bulk water, which is $\kappa_b=$\SI{5.5}{\micro\siemens/\meter}. Note that we neglected here so-called surface conductivity \cite{Bocquet2010}, due to the low reactivity of alumina with water at these pH \cite{wang2021investigations}. Moreover, this hypothesis is justified by the fact we observed a conductivity in the water film even lower than the water bulk conductivity, showing then a negligible contribution of added contributions such as the surface one.

The trend line indicates that conductivity increases with humidity but remains lower than that of bulk water for the RH values studied here. This result is remarkable, as conductivity is expected to be inversely proportional to $h_{RH}$. Wang et al.~\cite{WANG2022943} suggest that strong hydrogen bonds at the water/$\alpha$-alumina interface lead to ordered water structures, slowing down the translational and rotational dynamics of interfacial water compared to bulk water. As the thickness increases, the behavior and conductivity approach that of bulk water. A similar phenomenon was reported previously~\cite{BenIEEE1987}. Other works suggest that this stagnant layer is characterized by sub-diffusive motion~\cite{Sarfati-1021acsnano} due to the well-structured organization of water molecules near the surface, indicating an `ice-like' configuration~\cite{Asay2005, C3FD00096F}. In films thicker than four to five monolayers, the water may recover its bulk properties~\cite{doi:10.1021/la700893w}.

Based on our results showing lower conductivity compared to the bulk value, we propose an analogous interpretation, pointing that a fraction of the film condensed on nanopore walls may consist of a stagnant layer with `ice-like' behavior. We can estimate the thickness of the stagnant layer, $h_s$, by calculating the difference between the total thickness, $h_{RH}$, and the mobile portion, using the conductance values from the points (Figure~\ref{Figure4}). The mobile portion is assumed to have the same conductivity as bulk water, $\kappa_b$, and the experimentally obtained conductance, $G$  is attributed to this portion. By applying these values to Equation~\ref{kappa}, we determine $h_s$ by subtracting the mobile part from the total film,

\begin{equation}
    h_{s} = h_{RH} - \frac{GL}{\kappa_b 2\pi r_{np}N_p}.
\end{equation}

By averaging the stagnant thickness based on measurements shown in Figure~\ref{Figure4}c, we determine a stagnant thickness of $h_s = 0.97\pm 0.34$\SI{}{\nano\meter}. Since a water monolayer is typically about \SI{0.3}{\nano\meter} thick, suggesting approximately three stagnant monolayers at the interface. This value is in good agreement with literature data, suggesting that the range of influence of the alumina surface inducing an ordered molecular arrangement was observed to extend about \SI{1}{\nano\meter} from the surface~\cite{ArgyrisJCP2011}.

Similarly, for silica interfaces, three layers with an `ice-like' structure have been observed~\cite{Asay2005}, although a single layer of stagnant water was reported \cite{aymeric2023}. These differences between alumina and silica have been documented, the alumina surface interacting more with polar water molecules and then influencing ordering at larger scale in the liquid water \cite{wang2021investigations}.

Thus, the thickness of the stagnant layer $h_{s}$ measured in our study aligns with previously observed values at water/solid interfaces. Despite the increased adsorption force due to post-plasma treatment, which, as previously calculated, increases approximately tenfold according to the Hamaker constant, we do not observe a thicker layer $h_{s}$. Therefore, despite the stronger binding energy, it appears that the range of `ice-like' structure at the solid/liquid interface does not change significantly.

\section{Conclusions}

In this study, we probed how the current response of water films inside nanopores changes with variations in two factors related to water adsorption: the material's wettability and the relative humidity. One parameter, wettability, pertains to the material, while the other, relative humidity, relates to the environmental conditions. A nanoporous membrane made of oxidized aluminum, alumina, naturally hydrophilic, was used in the research and its wettability was tuned by plasma treatment. We noted an increase corresponding to higher relative humidity (RH) and wettability, implying a correlation with the amount of adsorbed water or pore filling thickness.

To quantify the amount of water adsorbed at different wettability levels, we used a balance to measure the quantity of water adsorbed by the membrane under varying relative humidity conditions. Applying Polanyi theory to describe the physics of this isotherm, we found that the adsorption energy, measured by the Hamaker constant, increased approximately tenfold after one minute of plasma cleaning compared to untreated cases. Regarding conductivity for the most hydrophilic case, which exhibited a stronger current response, we found that the values were lower than that of bulk distilled water ($\kappa_b$), approaching it as RH increases. This behavior is attributed to interactions with pore walls, which induce ordered stagnant layers near the interface. Our findings suggest the presence of approximately three stagnant monolayers. This value is in accordance with previous studies, indicating that the increase in adsorption energy does not lead to thicker immobile layers. Consequently, we can see that higher hydrophilicity enhances transport properties due to the increase in mobile layers. Our findings are important for understanding the properties of water films adsorbed by nanopores under different wettability conditions and environmental humidities, and  contribute to the development of future nanofluidic devices, at the scale of a few nanometers.

\begin{acknowledgement}

The authors thank Agnès Piednoir for AFM measurements and Rémy Fulcrand for SEM observations and deposition of Pt layer on membranes. A.-L. Biance thanks O. Vincent for fruitful discussions.   F.R. Leivas thanks the financial support given by the Eiffel fellowship program (project 118952W). This project was funded by the ANR project Soft Nanoflu (ANR-20-CE09-0025-03).
\end{acknowledgement}

\

\bibliography{main}

\providecommand{\latin}[1]{#1}
\makeatletter
\providecommand{\doi}
  {\begingroup\let\do\@makeother\dospecials
  \catcode`\{=1 \catcode`\}=2 \doi@aux}
\providecommand{\doi@aux}[1]{\endgroup\texttt{#1}}
\makeatother
\providecommand*\mcitethebibliography{\thebibliography}
\csname @ifundefined\endcsname{endmcitethebibliography}  {\let\endmcitethebibliography\endthebibliography}{}
\begin{mcitethebibliography}{88}
\providecommand*\natexlab[1]{#1}
\providecommand*\mciteSetBstSublistMode[1]{}
\providecommand*\mciteSetBstMaxWidthForm[2]{}
\providecommand*\mciteBstWouldAddEndPuncttrue
  {\def\EndOfBibitem{\unskip.}}
\providecommand*\mciteBstWouldAddEndPunctfalse
  {\let\EndOfBibitem\relax}
\providecommand*\mciteSetBstMidEndSepPunct[3]{}
\providecommand*\mciteSetBstSublistLabelBeginEnd[3]{}
\providecommand*\EndOfBibitem{}
\mciteSetBstSublistMode{f}
\mciteSetBstMaxWidthForm{subitem}{(\alph{mcitesubitemcount})}
\mciteSetBstSublistLabelBeginEnd
  {\mcitemaxwidthsubitemform\space}
  {\relax}
  {\relax}

\bibitem[Bonilla-Petriciolet \latin{et~al.}(2017)Bonilla-Petriciolet, Mendoza-Castillo, and Reynel-{\'A}vila]{bonilla2017adsorption}
Bonilla-Petriciolet,~A.; Mendoza-Castillo,~D.~I.; Reynel-{\'A}vila,~H.~E. \emph{Adsorption processes for water treatment and purification}; Springer, 2017; Vol. 256\relax
\mciteBstWouldAddEndPuncttrue
\mciteSetBstMidEndSepPunct{\mcitedefaultmidpunct}
{\mcitedefaultendpunct}{\mcitedefaultseppunct}\relax
\EndOfBibitem
\bibitem[Sarapat \latin{et~al.}(2020)Sarapat, Cox, and Baowan]{sarapat2020}
Sarapat,~N.,~P.and~Thamwattana; Cox,~B.~J.; Baowan,~D. Modelling carbon nanocones for selective filter. \emph{Journal of Mathematical Chemistry} \textbf{2020}, \emph{58}, 1650–1662\relax
\mciteBstWouldAddEndPuncttrue
\mciteSetBstMidEndSepPunct{\mcitedefaultmidpunct}
{\mcitedefaultendpunct}{\mcitedefaultseppunct}\relax
\EndOfBibitem
\bibitem[Kleinubing~Abal and Barbosa(2021)Kleinubing~Abal, and Barbosa]{Abal2021}
Kleinubing~Abal,~J.~P.; Barbosa,~M.~C. Molecular fluid flow in MoS2 nanoporous membranes and hydrodynamics interactions. \emph{The Journal of Chemical Physics} \textbf{2021}, \emph{154}, 134506\relax
\mciteBstWouldAddEndPuncttrue
\mciteSetBstMidEndSepPunct{\mcitedefaultmidpunct}
{\mcitedefaultendpunct}{\mcitedefaultseppunct}\relax
\EndOfBibitem
\bibitem[K\"{o}hler \latin{et~al.}(2018)K\"{o}hler, Bordin, {da Silva}, and Barbosa]{KOHLER2018331}
K\"{o}hler,~M.~H.; Bordin,~J.~R.; {da Silva},~L.~B.; Barbosa,~M.~C. Structure and dynamics of water inside hydrophobic and hydrophilic nanotubes. \emph{Physica A: Statistical Mechanics and its Applications} \textbf{2018}, \emph{490}, 331--337\relax
\mciteBstWouldAddEndPuncttrue
\mciteSetBstMidEndSepPunct{\mcitedefaultmidpunct}
{\mcitedefaultendpunct}{\mcitedefaultseppunct}\relax
\EndOfBibitem
\bibitem[Xu \latin{et~al.}(2014)Xu, Chen, Lu, Zhao, and Qiao]{Baoxing2014}
Xu,~B.; Chen,~X.; Lu,~W.; Zhao,~C.; Qiao,~Y. {Non-dissipative energy capture of confined liquid in nanopores}. \emph{Applied Physics Letters} \textbf{2014}, \emph{104}, 203107\relax
\mciteBstWouldAddEndPuncttrue
\mciteSetBstMidEndSepPunct{\mcitedefaultmidpunct}
{\mcitedefaultendpunct}{\mcitedefaultseppunct}\relax
\EndOfBibitem
\bibitem[Fernández \latin{et~al.}(2020)Fernández, Smith, and Johnson]{fernandez2020advances}
Fernández,~E.; Smith,~A.; Johnson,~R. Advances in harvesting water and energy from ubiquitous atmospheric moisture. \emph{Environmental Science \& Technology} \textbf{2020}, \emph{54}, 5863--5879\relax
\mciteBstWouldAddEndPuncttrue
\mciteSetBstMidEndSepPunct{\mcitedefaultmidpunct}
{\mcitedefaultendpunct}{\mcitedefaultseppunct}\relax
\EndOfBibitem
\bibitem[Brantley \latin{et~al.}(2008)Brantley, Kubicki, and White]{brantley2008kinetics}
Brantley,~S.~L.; Kubicki,~J.~D.; White,~A.~F. \emph{Kinetics of Water-Rock Interaction}; Springer: New York, 2008\relax
\mciteBstWouldAddEndPuncttrue
\mciteSetBstMidEndSepPunct{\mcitedefaultmidpunct}
{\mcitedefaultendpunct}{\mcitedefaultseppunct}\relax
\EndOfBibitem
\bibitem[Wheeler and Stroock(2008)Wheeler, and Stroock]{wheeler2008}
Wheeler,~T.~D.; Stroock,~A.~D. The transpiration of water at negative pressures in a synthetic tree. \emph{Nature} \textbf{2008}, \emph{455}, 208--212\relax
\mciteBstWouldAddEndPuncttrue
\mciteSetBstMidEndSepPunct{\mcitedefaultmidpunct}
{\mcitedefaultendpunct}{\mcitedefaultseppunct}\relax
\EndOfBibitem
\bibitem[Cira \latin{et~al.}(2015)Cira, Benusiglio, and Prakash]{Cira2015}
Cira,~N.; Benusiglio,~A.; Prakash,~M. Vapour-mediated sensing and motility in two-component droplets. \emph{Nature} \textbf{2015}, \emph{519}, 446—450\relax
\mciteBstWouldAddEndPuncttrue
\mciteSetBstMidEndSepPunct{\mcitedefaultmidpunct}
{\mcitedefaultendpunct}{\mcitedefaultseppunct}\relax
\EndOfBibitem
\bibitem[Gyurcsányi(2008)]{GYURCSANYI2008627}
Gyurcsányi,~R.~E. Chemically-modified nanopores for sensing. \emph{TrAC Trends in Analytical Chemistry} \textbf{2008}, \emph{27}, 627--639, Electroanalysis Based on Nanomaterials\relax
\mciteBstWouldAddEndPuncttrue
\mciteSetBstMidEndSepPunct{\mcitedefaultmidpunct}
{\mcitedefaultendpunct}{\mcitedefaultseppunct}\relax
\EndOfBibitem
\bibitem[Wang \latin{et~al.}(2013)Wang, Wang, Han, Zhou, and Guan]{Wang2013}
Wang,~G.; Wang,~L.; Han,~Y.; Zhou,~S.; Guan,~X. Nanopore Stochastic Detection: Diversity, Sensitivity, and Beyond. \emph{Accounts of Chemical Research} \textbf{2013}, \emph{46}, 2867--2877, PMID: 23614724\relax
\mciteBstWouldAddEndPuncttrue
\mciteSetBstMidEndSepPunct{\mcitedefaultmidpunct}
{\mcitedefaultendpunct}{\mcitedefaultseppunct}\relax
\EndOfBibitem
\bibitem[Bocquet and Charlaix(2010)Bocquet, and Charlaix]{Bocquet2010}
Bocquet,~L.; Charlaix,~E. Nanofluidics{,} from bulk to interfaces. \emph{Chem. Soc. Rev.} \textbf{2010}, \emph{39}, 1073--1095\relax
\mciteBstWouldAddEndPuncttrue
\mciteSetBstMidEndSepPunct{\mcitedefaultmidpunct}
{\mcitedefaultendpunct}{\mcitedefaultseppunct}\relax
\EndOfBibitem
\bibitem[Rykaczewski \latin{et~al.}(2012)Rykaczewski, Osborn, Chinn, Walker, Scott, Jones, Hao, Yao, and Wang]{C2SM25502B}
Rykaczewski,~K.; Osborn,~W.~A.; Chinn,~J.; Walker,~M.~L.; Scott,~J. H.~J.; Jones,~W.; Hao,~C.; Yao,~S.; Wang,~Z. How nanorough is rough enough to make a surface superhydrophobic during water condensation? \emph{Soft Matter} \textbf{2012}, \emph{8}, 8786--8794\relax
\mciteBstWouldAddEndPuncttrue
\mciteSetBstMidEndSepPunct{\mcitedefaultmidpunct}
{\mcitedefaultendpunct}{\mcitedefaultseppunct}\relax
\EndOfBibitem
\bibitem[Ranathunga \latin{et~al.}(2020)Ranathunga, Shamir, Dai, and Nielsen]{Ranathunga2020}
Ranathunga,~D. T.~S.; Shamir,~A.; Dai,~X.; Nielsen,~S.~O. Molecular Dynamics Simulations of Water Condensation on Surfaces with Tunable Wettability. \emph{Langmuir} \textbf{2020}, \emph{36}, 7383--7391, PMID: 32498521\relax
\mciteBstWouldAddEndPuncttrue
\mciteSetBstMidEndSepPunct{\mcitedefaultmidpunct}
{\mcitedefaultendpunct}{\mcitedefaultseppunct}\relax
\EndOfBibitem
\bibitem[Edalatpour \latin{et~al.}(2018)Edalatpour, Liu, Jacobi, Eid, and Sommers]{EDALATPOUR2018967}
Edalatpour,~M.; Liu,~L.; Jacobi,~A.; Eid,~K.; Sommers,~A. Managing water on heat transfer surfaces: A critical review of techniques to modify surface wettability for applications with condensation or evaporation. \emph{Applied Energy} \textbf{2018}, \emph{222}, 967--992\relax
\mciteBstWouldAddEndPuncttrue
\mciteSetBstMidEndSepPunct{\mcitedefaultmidpunct}
{\mcitedefaultendpunct}{\mcitedefaultseppunct}\relax
\EndOfBibitem
\bibitem[Seki \latin{et~al.}(2021)Seki, Takamatsu, Suzuki, Oya, and Ohba]{ohba2021}
Seki,~R.; Takamatsu,~H.; Suzuki,~Y.; Oya,~Y.; Ohba,~T. Hydrophobic-to-hydrophilic affinity change of sub-monolayer water molecules at water–graphene interfaces. \emph{Colloids and Surfaces A: Physicochemical and Engineering Aspects} \textbf{2021}, \emph{628}, 127393\relax
\mciteBstWouldAddEndPuncttrue
\mciteSetBstMidEndSepPunct{\mcitedefaultmidpunct}
{\mcitedefaultendpunct}{\mcitedefaultseppunct}\relax
\EndOfBibitem
\bibitem[Köhler \latin{et~al.}(2019)Köhler, Bordin, {de Matos}, and Barbosa]{kohler2019}
Köhler,~M.~H.; Bordin,~J.~R.; {de Matos},~C.~F.; Barbosa,~M.~C. Water in nanotubes: The surface effect. \emph{Chemical Engineering Science} \textbf{2019}, \emph{203}, 54--67\relax
\mciteBstWouldAddEndPuncttrue
\mciteSetBstMidEndSepPunct{\mcitedefaultmidpunct}
{\mcitedefaultendpunct}{\mcitedefaultseppunct}\relax
\EndOfBibitem
\bibitem[Shaat and Zheng(2019)Shaat, and Zheng]{zheng2019}
Shaat,~M.; Zheng,~Y. Fluidity and phase transitions of water in hydrophobic and hydrophilic nanotubes. \emph{Scientific Reports} \textbf{2019}, \emph{9}, 5689\relax
\mciteBstWouldAddEndPuncttrue
\mciteSetBstMidEndSepPunct{\mcitedefaultmidpunct}
{\mcitedefaultendpunct}{\mcitedefaultseppunct}\relax
\EndOfBibitem
\bibitem[R.Leivas and Barbosa(2023)R.Leivas, and Barbosa]{Leivas_JCP}
R.Leivas,~F.; Barbosa,~M.~C. {Functionalized carbon nanocones performance in water harvesting}. \emph{The Journal of Chemical Physics} \textbf{2023}, \emph{158}, 194702\relax
\mciteBstWouldAddEndPuncttrue
\mciteSetBstMidEndSepPunct{\mcitedefaultmidpunct}
{\mcitedefaultendpunct}{\mcitedefaultseppunct}\relax
\EndOfBibitem
\bibitem[Leivas and Barbosa(2023)Leivas, and Barbosa]{Leivas_Beils}
Leivas,~F.~R.; Barbosa,~M.~C. Atmospheric water harvesting using functionalized carbon nanocones. \emph{Beilstein Journal of Nanotechnology} \textbf{2023}, \emph{14}, 1--10\relax
\mciteBstWouldAddEndPuncttrue
\mciteSetBstMidEndSepPunct{\mcitedefaultmidpunct}
{\mcitedefaultendpunct}{\mcitedefaultseppunct}\relax
\EndOfBibitem
\bibitem[Jiang \latin{et~al.}(2022)Jiang, Song, Jia, Yang, Li, Li, and Du]{JIANG20223437}
Jiang,~W.; Song,~J.; Jia,~T.; Yang,~L.; Li,~S.; Li,~Y.; Du,~K. A comprehensive review on the pre-research of nanofluids in absorption refrigeration systems. \emph{Energy Reports} \textbf{2022}, \emph{8}, 3437--3464\relax
\mciteBstWouldAddEndPuncttrue
\mciteSetBstMidEndSepPunct{\mcitedefaultmidpunct}
{\mcitedefaultendpunct}{\mcitedefaultseppunct}\relax
\EndOfBibitem
\bibitem[Weerakoon-Ratnayake \latin{et~al.}(2019)Weerakoon-Ratnayake, Vaidyanathan, Amarasekara, Johnson, and Soper]{WEERAKOON2019335}
Weerakoon-Ratnayake,~K.~M.; Vaidyanathan,~S.; Amarasekara,~C.~A.; Johnson,~C.~K.; Soper,~S.~A. In \emph{Spectroscopy and Dynamics of Single Molecules}; Johnson,~C.~K., Ed.; Developments in Physical \& Theoretical Chemistry; Elsevier, 2019; pp 335--377\relax
\mciteBstWouldAddEndPuncttrue
\mciteSetBstMidEndSepPunct{\mcitedefaultmidpunct}
{\mcitedefaultendpunct}{\mcitedefaultseppunct}\relax
\EndOfBibitem
\bibitem[Eijkel and Berg(2005)Eijkel, and Berg]{eijkel2005nanofluidics}
Eijkel,~J. C.~T.; Berg,~A. v.~d. Nanofluidics: What is it and what can we expect from it? \emph{Microfluidics and Nanofluidics} \textbf{2005}, \emph{1}, 249--267\relax
\mciteBstWouldAddEndPuncttrue
\mciteSetBstMidEndSepPunct{\mcitedefaultmidpunct}
{\mcitedefaultendpunct}{\mcitedefaultseppunct}\relax
\EndOfBibitem
\bibitem[Tinti \latin{et~al.}(2021)Tinti, Camisasca, and Giacomello]{giacomello2021}
Tinti,~A.; Camisasca,~G.; Giacomello,~A. Structure and dynamics of water confined in cylindrical nanopores with varying hydrophobicity. \emph{Philosophical Transactions of the Royal Society A: Mathematical, Physical and Engineering Sciences} \textbf{2021}, \emph{379}, 20200403\relax
\mciteBstWouldAddEndPuncttrue
\mciteSetBstMidEndSepPunct{\mcitedefaultmidpunct}
{\mcitedefaultendpunct}{\mcitedefaultseppunct}\relax
\EndOfBibitem
\bibitem[Mondal and Bagchi(2020)Mondal, and Bagchi]{bagchi2020}
Mondal,~S.; Bagchi,~B. How different are the dynamics of nanoconfined water? \emph{Journal of Chemical Physics} \textbf{2020}, \emph{152}, 224707\relax
\mciteBstWouldAddEndPuncttrue
\mciteSetBstMidEndSepPunct{\mcitedefaultmidpunct}
{\mcitedefaultendpunct}{\mcitedefaultseppunct}\relax
\EndOfBibitem
\bibitem[Parmentier \latin{et~al.}(2019)Parmentier, Maccarini, Francesco, Scaccia, Rogati, Czakkel, and Luca]{luca2019}
Parmentier,~A.; Maccarini,~M.; Francesco,~A.~D.; Scaccia,~L.; Rogati,~G.; Czakkel,~O.; Luca,~F.~D. Neutron spin echo monitoring of segmental-like diffusion of water confined in the cores of carbon nanotubes. \emph{Phys. Chem. Chem. Phys.} \textbf{2019}, \emph{21}, 21456\relax
\mciteBstWouldAddEndPuncttrue
\mciteSetBstMidEndSepPunct{\mcitedefaultmidpunct}
{\mcitedefaultendpunct}{\mcitedefaultseppunct}\relax
\EndOfBibitem
\bibitem[Wu \latin{et~al.}(2017)Wu, Chen, Li, Li, Xu, and Dong]{donga2017}
Wu,~K.; Chen,~Z.; Li,~J.; Li,~X.; Xu,~J.; Dong,~X. Wettability effect on nanoconfined water flow. \emph{Proceedings of the National Academy of Sciences} \textbf{2017}, \emph{114}, 3358--3363\relax
\mciteBstWouldAddEndPuncttrue
\mciteSetBstMidEndSepPunct{\mcitedefaultmidpunct}
{\mcitedefaultendpunct}{\mcitedefaultseppunct}\relax
\EndOfBibitem
\bibitem[Qi \latin{et~al.}(2023)Qi, Ling, and Wang]{Chonghai2023}
Qi,~C.; Ling,~C.; Wang,~C. Ordered/Disordered Structures of Water at Solid/Liquid Interfaces. \emph{Crystals} \textbf{2023}, \emph{13}, 263\relax
\mciteBstWouldAddEndPuncttrue
\mciteSetBstMidEndSepPunct{\mcitedefaultmidpunct}
{\mcitedefaultendpunct}{\mcitedefaultseppunct}\relax
\EndOfBibitem
\bibitem[Chen \latin{et~al.}(2013)Chen, Gao, Wang, Zhao, and Fang]{Chen2013}
Chen,~J.; Gao,~Y.; Wang,~C.; Zhao,~H.; Fang,~H. Ice or water: Thermal properties of monolayer water adsorbed on a substrate. \emph{Journal of Statistical Mechanics: Theory and Experiment} \textbf{2013}, \emph{2013}, P06009\relax
\mciteBstWouldAddEndPuncttrue
\mciteSetBstMidEndSepPunct{\mcitedefaultmidpunct}
{\mcitedefaultendpunct}{\mcitedefaultseppunct}\relax
\EndOfBibitem
\bibitem[Xiao \latin{et~al.}(2019)Xiao, Shi, Yan, Chen, Qian, and Kim]{colloids3030055}
Xiao,~C.; Shi,~P.; Yan,~W.; Chen,~L.; Qian,~L.; Kim,~S.~H. Thickness and Structure of Adsorbed Water Layer and Effects on Adhesion and Friction at Nanoasperity Contact. \emph{Colloids and Interfaces} \textbf{2019}, \emph{3}\relax
\mciteBstWouldAddEndPuncttrue
\mciteSetBstMidEndSepPunct{\mcitedefaultmidpunct}
{\mcitedefaultendpunct}{\mcitedefaultseppunct}\relax
\EndOfBibitem
\bibitem[Dette \latin{et~al.}(2018)Dette, Pérez-Osorio, Mangel, Giustino, Jung, and Kern]{Dette2018}
Dette,~C.; Pérez-Osorio,~M.~A.; Mangel,~S.; Giustino,~F.; Jung,~S.~J.; Kern,~K. Atomic Structure of Water Monolayer on Anatase TiO2(101) Surface. \emph{The Journal of Physical Chemistry C} \textbf{2018}, \emph{122}, 11954--11960\relax
\mciteBstWouldAddEndPuncttrue
\mciteSetBstMidEndSepPunct{\mcitedefaultmidpunct}
{\mcitedefaultendpunct}{\mcitedefaultseppunct}\relax
\EndOfBibitem
\bibitem[Meng \latin{et~al.}(2004)Meng, Wang, and Gao]{PhysRevB.69.195404}
Meng,~S.; Wang,~E.~G.; Gao,~S. Water adsorption on metal surfaces: A general picture from density functional theory studies. \emph{Phys. Rev. B} \textbf{2004}, \emph{69}, 195404\relax
\mciteBstWouldAddEndPuncttrue
\mciteSetBstMidEndSepPunct{\mcitedefaultmidpunct}
{\mcitedefaultendpunct}{\mcitedefaultseppunct}\relax
\EndOfBibitem
\bibitem[Algara-Siller \latin{et~al.}(2015)Algara-Siller, Lehtinen, Wang, Nair, Kaiser, Wu, Geim, and Grigorieva]{algara2015square}
Algara-Siller,~G.; Lehtinen,~O.; Wang,~F.; Nair,~R.~R.; Kaiser,~U.; Wu,~H.; Geim,~A.~K.; Grigorieva,~I.~V. Square ice in graphene nanocapillaries. \emph{Nature} \textbf{2015}, \emph{519}, 443--445\relax
\mciteBstWouldAddEndPuncttrue
\mciteSetBstMidEndSepPunct{\mcitedefaultmidpunct}
{\mcitedefaultendpunct}{\mcitedefaultseppunct}\relax
\EndOfBibitem
\bibitem[Allemand \latin{et~al.}(2023)Allemand, Zhao, Vincent, Fulcrand, Joly, Ybert, and Biance]{aymeric2023}
Allemand,~A.; Zhao,~M.; Vincent,~O.; Fulcrand,~R.; Joly,~L.; Ybert,~C.; Biance,~A.-L. Anomalous ionic transport in tunable angstrom-size water films on silica. \emph{Proceedings of the National Academy of Sciences} \textbf{2023}, \emph{120}, e2221304120\relax
\mciteBstWouldAddEndPuncttrue
\mciteSetBstMidEndSepPunct{\mcitedefaultmidpunct}
{\mcitedefaultendpunct}{\mcitedefaultseppunct}\relax
\EndOfBibitem
\bibitem[Younssi \latin{et~al.}(2018)Younssi, Breida, and Achiou]{younssi2018alumina}
Younssi,~S.~A.; Breida,~M.; Achiou,~B. Alumina Membranes for Desalination and Water Treatment. \emph{Desalination and Water Treatment} \textbf{2018}, \relax
\mciteBstWouldAddEndPunctfalse
\mciteSetBstMidEndSepPunct{\mcitedefaultmidpunct}
{}{\mcitedefaultseppunct}\relax
\EndOfBibitem
\bibitem[Ben-Nissan \latin{et~al.}(2008)Ben-Nissan, Choi, and Cordingley]{BENNISSAN2008223}
Ben-Nissan,~B.; Choi,~A.~H.; Cordingley,~R. In \emph{Bioceramics and their Clinical Applications}; Kokubo,~T., Ed.; Woodhead Publishing Series in Biomaterials; Woodhead Publishing, 2008; pp 223--242\relax
\mciteBstWouldAddEndPuncttrue
\mciteSetBstMidEndSepPunct{\mcitedefaultmidpunct}
{\mcitedefaultendpunct}{\mcitedefaultseppunct}\relax
\EndOfBibitem
\bibitem[Wang \latin{et~al.}(2022)Wang, Ma, Ulbricht, Dong, and Zhao]{wang2022progress}
Wang,~Y.; Ma,~B.; Ulbricht,~M.; Dong,~Y.; Zhao,~X. Progress in alumina ceramic membranes for water purification: Status and prospects. \emph{Water Research} \textbf{2022}, \emph{226}, 119173\relax
\mciteBstWouldAddEndPuncttrue
\mciteSetBstMidEndSepPunct{\mcitedefaultmidpunct}
{\mcitedefaultendpunct}{\mcitedefaultseppunct}\relax
\EndOfBibitem
\bibitem[Li \latin{et~al.}(2018)Li, Zhao, Liu, Zhang, Zhang, and Xie]{li2018highly}
Li,~W.; Zhao,~Y.; Liu,~C.; Zhang,~H.; Zhang,~L.; Xie,~Y. Highly stable and self-cleaning ultrafiltration membranes from aluminum oxide nanoparticles for drinking water treatment. \emph{Journal of Membrane Science} \textbf{2018}, \emph{562}, 60--69\relax
\mciteBstWouldAddEndPuncttrue
\mciteSetBstMidEndSepPunct{\mcitedefaultmidpunct}
{\mcitedefaultendpunct}{\mcitedefaultseppunct}\relax
\EndOfBibitem
\bibitem[Yang \latin{et~al.}(2020)Yang, Huang, and Men]{yang2020alumina}
Yang,~J.; Huang,~A.; Men,~L. Alumina membranes for water treatment: preparation, characterization, and application. \emph{Reviews in Chemical Engineering} \textbf{2020}, \relax
\mciteBstWouldAddEndPunctfalse
\mciteSetBstMidEndSepPunct{\mcitedefaultmidpunct}
{}{\mcitedefaultseppunct}\relax
\EndOfBibitem
\bibitem[Chen \latin{et~al.}(2013)Chen, Li, Li, and Zhu]{chen2013high}
Chen,~D.; Li,~Q.; Li,~J.; Zhu,~X. High-K Dielectric Properties of Alumina Ceramics for Electronic Applications. \emph{Journal of the American Ceramic Society} \textbf{2013}, \emph{96}, 373--378\relax
\mciteBstWouldAddEndPuncttrue
\mciteSetBstMidEndSepPunct{\mcitedefaultmidpunct}
{\mcitedefaultendpunct}{\mcitedefaultseppunct}\relax
\EndOfBibitem
\bibitem[Gavazzoni \latin{et~al.}(2017)Gavazzoni, Giovambattista, Netz, and Barbosa]{gavazzoni2017}
Gavazzoni,~C.; Giovambattista,~N.; Netz,~P.~A.; Barbosa,~M.~C. Structure and mobility of water confined in AlPO4-54 nanotubes. \emph{Journal of Chemical Physics} \textbf{2017}, \emph{146}, 234509\relax
\mciteBstWouldAddEndPuncttrue
\mciteSetBstMidEndSepPunct{\mcitedefaultmidpunct}
{\mcitedefaultendpunct}{\mcitedefaultseppunct}\relax
\EndOfBibitem
\bibitem[Zhang \latin{et~al.}(2008)Zhang, Tian, Waychunas, and Shen]{ZhangJACS2008}
Zhang,~L.; Tian,~C.; Waychunas,~G.~A.; Shen,~Y.~R. Structures and Charging of $\alpha$-Alumina (0001)/Water Interfaces Studied by Sum-Frequency Vibrational Spectroscopy. \emph{Journal of the American Chemical Society} \textbf{2008}, \emph{130}, 7686--7694, PMID: 18491896\relax
\mciteBstWouldAddEndPuncttrue
\mciteSetBstMidEndSepPunct{\mcitedefaultmidpunct}
{\mcitedefaultendpunct}{\mcitedefaultseppunct}\relax
\EndOfBibitem
\bibitem[Franks and Gan()Franks, and Gan]{Franks2007}
Franks,~G.~V.; Gan,~Y. Charging Behavior at the Alumina–Water Interface and Implications for Ceramic Processing. \emph{Journal of the American Ceramic Society} \emph{90}, 3373--3388\relax
\mciteBstWouldAddEndPuncttrue
\mciteSetBstMidEndSepPunct{\mcitedefaultmidpunct}
{\mcitedefaultendpunct}{\mcitedefaultseppunct}\relax
\EndOfBibitem
\bibitem[Petrik \latin{et~al.}(2018)Petrik, Huestis, LaVerne, Aleksandrov, Orlando, and Kimmel]{PetrikJCP2018}
Petrik,~N.~G.; Huestis,~P.~L.; LaVerne,~J.~A.; Aleksandrov,~A.~B.; Orlando,~T.~M.; Kimmel,~G.~A. Molecular Water Adsorption and Reactions on $\alpha$-Al2O3(0001) and $\alpha$-Alumina Particles. \emph{The Journal of Physical Chemistry C} \textbf{2018}, \emph{122}, 9540--9551\relax
\mciteBstWouldAddEndPuncttrue
\mciteSetBstMidEndSepPunct{\mcitedefaultmidpunct}
{\mcitedefaultendpunct}{\mcitedefaultseppunct}\relax
\EndOfBibitem
\bibitem[Kirsch \latin{et~al.}(2014)Kirsch, Wirth, Tong, Wolf, Saalfrank, and Campen]{Kirsch2014}
Kirsch,~H.; Wirth,~J.; Tong,~Y.; Wolf,~M.; Saalfrank,~P.; Campen,~R.~K. Experimental Characterization of Unimolecular Water Dissociative Adsorption on $\alpha$-Alumina. \emph{The Journal of Physical Chemistry C} \textbf{2014}, \emph{118}, 13623--13630\relax
\mciteBstWouldAddEndPuncttrue
\mciteSetBstMidEndSepPunct{\mcitedefaultmidpunct}
{\mcitedefaultendpunct}{\mcitedefaultseppunct}\relax
\EndOfBibitem
\bibitem[Tong \latin{et~al.}(2015)Tong, Wirth, Kirsch, Wolf, Saalfrank, and Campen]{Tong2015}
Tong,~Y.; Wirth,~J.; Kirsch,~H.; Wolf,~M.; Saalfrank,~P.; Campen,~R. Optically probing Al—O and O—H vibrations to characterize water adsorption and surface reconstruction on $\alpha$-alumina: An experimental and theoretical study. \emph{The Journal of Chemical Physics} \textbf{2015}, \emph{142}, 54704\relax
\mciteBstWouldAddEndPuncttrue
\mciteSetBstMidEndSepPunct{\mcitedefaultmidpunct}
{\mcitedefaultendpunct}{\mcitedefaultseppunct}\relax
\EndOfBibitem
\bibitem[Elam \latin{et~al.}(1998)Elam, Nelson, Cameron, Tolbert, and George]{Elam1998}
Elam,~J.~W.; Nelson,~C.~E.; Cameron,~M.~A.; Tolbert,~M.~A.; George,~S.~M. Adsorption of H2O on a Single-Crystal $\alpha$-Al2O3(0001) Surface. \emph{The Journal of Physical Chemistry B} \textbf{1998}, \emph{102}, 7008--7015\relax
\mciteBstWouldAddEndPuncttrue
\mciteSetBstMidEndSepPunct{\mcitedefaultmidpunct}
{\mcitedefaultendpunct}{\mcitedefaultseppunct}\relax
\EndOfBibitem
\bibitem[Nelson \latin{et~al.}(1998)Nelson, Elam, Cameron, Tolbert, and George]{NELSON1998341}
Nelson,~C.; Elam,~J.; Cameron,~M.; Tolbert,~M.; George,~S. Desorption of H2O from a hydroxylated single-crystal $\alpha$-Al2O3(0001) surface. \emph{Surface Science} \textbf{1998}, \emph{416}, 341--353\relax
\mciteBstWouldAddEndPuncttrue
\mciteSetBstMidEndSepPunct{\mcitedefaultmidpunct}
{\mcitedefaultendpunct}{\mcitedefaultseppunct}\relax
\EndOfBibitem
\bibitem[Prasetyo and Hofer()Prasetyo, and Hofer]{PRASETYO2019195}
Prasetyo,~N.; Hofer,~T.~S. Adsorption and dissociation of water molecules at the $\alpha$-Al2O3(0001) surface: A 2-dimensional hybrid self-consistent charge density functional based tight-binding/molecular mechanics molecular dynamics (2D SCC-DFTB/MM MD) simulation study. \emph{Computational Materials Science} \emph{164}, 195--204\relax
\mciteBstWouldAddEndPuncttrue
\mciteSetBstMidEndSepPunct{\mcitedefaultmidpunct}
{\mcitedefaultendpunct}{\mcitedefaultseppunct}\relax
\EndOfBibitem
\bibitem[Yue \latin{et~al.}(2022)Yue, Melani, Kirsch, Paarmann, Saalfrank, Campen, and Tong]{Yanhua2022}
Yue,~Y.; Melani,~G.; Kirsch,~H.; Paarmann,~A.; Saalfrank,~P.; Campen,~R.~K.; Tong,~Y. Structure and Reactivity of $\alpha$-Al2O3(0001) Surfaces: How Do Al–I and Gibbsite-like Terminations Interconvert? \emph{The Journal of Physical Chemistry C} \textbf{2022}, \emph{126}, 13467--13476\relax
\mciteBstWouldAddEndPuncttrue
\mciteSetBstMidEndSepPunct{\mcitedefaultmidpunct}
{\mcitedefaultendpunct}{\mcitedefaultseppunct}\relax
\EndOfBibitem
\bibitem[Lin \latin{et~al.}(2020)Lin, Chuang, Hsiao, Yeh, and Ho]{Kuo-Chuan2020}
Lin,~Y.-C.; Chuang,~C.-H.; Hsiao,~L.-Y.; Yeh,~M.-H.; Ho,~K.-C. Oxygen Plasma Activation of Carbon Nanotubes-Interconnected Prussian Blue Analogue for Oxygen Evolution Reaction. \emph{ACS Applied Materials \& Interfaces} \textbf{2020}, \emph{12}, 42634--42643, PMID: 32845608\relax
\mciteBstWouldAddEndPuncttrue
\mciteSetBstMidEndSepPunct{\mcitedefaultmidpunct}
{\mcitedefaultendpunct}{\mcitedefaultseppunct}\relax
\EndOfBibitem
\bibitem[Yu \latin{et~al.}(2018)Yu, Li, Zhang, Li, Dong, Zhang, and Huang]{YU2018383}
Yu,~S.; Li,~J.; Zhang,~Y.; Li,~M.; Dong,~F.; Zhang,~T.; Huang,~H. Local spatial charge separation and proton activation induced by surface hydroxylation promoting photocatalytic hydrogen evolution of polymeric carbon nitride. \emph{Nano Energy} \textbf{2018}, \emph{50}, 383--392\relax
\mciteBstWouldAddEndPuncttrue
\mciteSetBstMidEndSepPunct{\mcitedefaultmidpunct}
{\mcitedefaultendpunct}{\mcitedefaultseppunct}\relax
\EndOfBibitem
\bibitem[Mehta \latin{et~al.}(2019)Mehta, Barboun, Go, Hicks, and Schneider]{WilliamF2019}
Mehta,~P.; Barboun,~P.; Go,~D.~B.; Hicks,~J.~C.; Schneider,~W.~F. Catalysis Enabled by Plasma Activation of Strong Chemical Bonds: A Review. \emph{ACS Energy Letters} \textbf{2019}, \emph{4}, 1115--1133\relax
\mciteBstWouldAddEndPuncttrue
\mciteSetBstMidEndSepPunct{\mcitedefaultmidpunct}
{\mcitedefaultendpunct}{\mcitedefaultseppunct}\relax
\EndOfBibitem
\bibitem[Kim \latin{et~al.}(2015)Kim, Sukovich, and Abate]{kim2015patterning}
Kim,~S.~C.; Sukovich,~D.~J.; Abate,~A.~R. Patterning microfluidic device wettability with spatially-controlled plasma oxidation. \emph{Lab on a Chip} \textbf{2015}, \emph{15}, 3163--3169\relax
\mciteBstWouldAddEndPuncttrue
\mciteSetBstMidEndSepPunct{\mcitedefaultmidpunct}
{\mcitedefaultendpunct}{\mcitedefaultseppunct}\relax
\EndOfBibitem
\bibitem[Sui \latin{et~al.}(2021)Sui, Li, Shen, Ni, Xie, Lin, Zhao, Guo, and Duan]{Siyuan2021}
Sui,~S.; Li,~L.; Shen,~J.; Ni,~G.; Xie,~H.; Lin,~Q.; Zhao,~Y.; Guo,~J.; Duan,~W. Plasma treatment of polymethyl methacrylate to improve surface hydrophilicity and antifouling performance. \emph{Polymer Engineering \& Science} \textbf{2021}, \emph{61}, 506--513\relax
\mciteBstWouldAddEndPuncttrue
\mciteSetBstMidEndSepPunct{\mcitedefaultmidpunct}
{\mcitedefaultendpunct}{\mcitedefaultseppunct}\relax
\EndOfBibitem
\bibitem[Wang \latin{et~al.}(2022)Wang, Zou, Remsing, Ross, Klein, Carnevale, and Borguet]{WANG2022943}
Wang,~R.; Zou,~Y.; Remsing,~R.~C.; Ross,~N.~O.; Klein,~M.~L.; Carnevale,~V.; Borguet,~E. Superhydrophilicity of $\alpha$-alumina surfaces results from tight binding of interfacial waters to specific aluminols. \emph{Journal of Colloid and Interface Science} \textbf{2022}, \emph{628}, 943--954\relax
\mciteBstWouldAddEndPuncttrue
\mciteSetBstMidEndSepPunct{\mcitedefaultmidpunct}
{\mcitedefaultendpunct}{\mcitedefaultseppunct}\relax
\EndOfBibitem
\bibitem[Polanyi(1916)]{polanyi1916adsorption}
Polanyi,~M. Adsorption of gases (vapors) by a solid non-volatile adsorbent. \emph{Verh. Dtsch. Phys. Ges} \textbf{1916}, \emph{18}, 55--80\relax
\mciteBstWouldAddEndPuncttrue
\mciteSetBstMidEndSepPunct{\mcitedefaultmidpunct}
{\mcitedefaultendpunct}{\mcitedefaultseppunct}\relax
\EndOfBibitem
\bibitem[Hu \latin{et~al.}(2019)Hu, Li, Liu, and Fang]{HU2019249}
Hu,~H.; Li,~Q.; Liu,~S.; Fang,~T. Molecular dynamics study on water vapor condensation and infiltration characteristics in nanopores with tunable wettability. \emph{Applied Surface Science} \textbf{2019}, \emph{494}, 249--258\relax
\mciteBstWouldAddEndPuncttrue
\mciteSetBstMidEndSepPunct{\mcitedefaultmidpunct}
{\mcitedefaultendpunct}{\mcitedefaultseppunct}\relax
\EndOfBibitem
\bibitem[Cihan \latin{et~al.}(2019)Cihan, Tokunaga, and Birkholzer]{CihanA2019}
Cihan,~A.; Tokunaga,~T.~K.; Birkholzer,~J.~T. Adsorption and Capillary Condensation-Induced Imbibition in Nanoporous Media. \emph{Langmuir} \textbf{2019}, \emph{35}, 9611--9621, PMID: 31241970\relax
\mciteBstWouldAddEndPuncttrue
\mciteSetBstMidEndSepPunct{\mcitedefaultmidpunct}
{\mcitedefaultendpunct}{\mcitedefaultseppunct}\relax
\EndOfBibitem
\bibitem[Urteaga \latin{et~al.}(2019)Urteaga, Mercuri, Gimenez, Bellino, and Berli]{URTEAGA2019407}
Urteaga,~R.; Mercuri,~M.; Gimenez,~R.; Bellino,~M.~G.; Berli,~C.~L. Spontaneous water adsorption-desorption oscillations in mesoporous thin films. \emph{Journal of Colloid and Interface Science} \textbf{2019}, \emph{537}, 407--413\relax
\mciteBstWouldAddEndPuncttrue
\mciteSetBstMidEndSepPunct{\mcitedefaultmidpunct}
{\mcitedefaultendpunct}{\mcitedefaultseppunct}\relax
\EndOfBibitem
\bibitem[van~den Berg \latin{et~al.}(2010)van~den Berg, Craighead, and Yang]{C001349H}
van~den Berg,~A.; Craighead,~H.~G.; Yang,~P. From microfluidic applications to nanofluidic phenomena. \emph{Chem. Soc. Rev.} \textbf{2010}, \emph{39}, 899--900\relax
\mciteBstWouldAddEndPuncttrue
\mciteSetBstMidEndSepPunct{\mcitedefaultmidpunct}
{\mcitedefaultendpunct}{\mcitedefaultseppunct}\relax
\EndOfBibitem
\bibitem[Xu \latin{et~al.}(2020)Xu, Wei, Zhang, and Wang]{XU2020118297}
Xu,~F.; Wei,~M.; Zhang,~X.; Wang,~Y. Effect of hydrophilicity on water transport through sub-nanometer pores. \emph{Journal of Membrane Science} \textbf{2020}, \emph{611}, 118297\relax
\mciteBstWouldAddEndPuncttrue
\mciteSetBstMidEndSepPunct{\mcitedefaultmidpunct}
{\mcitedefaultendpunct}{\mcitedefaultseppunct}\relax
\EndOfBibitem
\bibitem[Pippard(1964)]{pippard1964elements}
Pippard,~A. \emph{Elements of Classical Thermodynamics:For Advanced Students of Physics}; Elements of Classical Thermodynamics for Advanced Students of Physics; Cambridge University Press, 1964\relax
\mciteBstWouldAddEndPuncttrue
\mciteSetBstMidEndSepPunct{\mcitedefaultmidpunct}
{\mcitedefaultendpunct}{\mcitedefaultseppunct}\relax
\EndOfBibitem
\bibitem[Wagner and Pruß(2002)Wagner, and Pruß]{Wagner2002}
Wagner,~W.; Pruß,~A. {The IAPWS Formulation 1995 for the Thermodynamic Properties of Ordinary Water Substance for General and Scientific Use}. \emph{Journal of Physical and Chemical Reference Data} \textbf{2002}, \emph{31}, 387--535\relax
\mciteBstWouldAddEndPuncttrue
\mciteSetBstMidEndSepPunct{\mcitedefaultmidpunct}
{\mcitedefaultendpunct}{\mcitedefaultseppunct}\relax
\EndOfBibitem
\bibitem[Isabell \latin{et~al.}(1999)Isabell, Fischione, O’Keefe, Guruz, and Dravid]{Isabell1999}
Isabell,~T.~C.; Fischione,~P.~E.; O’Keefe,~C.; Guruz,~M.~U.; Dravid,~V.~P. Plasma Cleaning and Its Applications for Electron Microscopy. \emph{Microscopy and Microanalysis} \textbf{1999}, \emph{5}, 126–135\relax
\mciteBstWouldAddEndPuncttrue
\mciteSetBstMidEndSepPunct{\mcitedefaultmidpunct}
{\mcitedefaultendpunct}{\mcitedefaultseppunct}\relax
\EndOfBibitem
\bibitem[Banerjee \latin{et~al.}(2010)Banerjee, Kumar, Bremmell, and Griesser]{banerjee2010molecular}
Banerjee,~K.~K.; Kumar,~S.; Bremmell,~K.~E.; Griesser,~H.~J. Molecular-level removal of proteinaceous contamination from model surfaces and biomedical device materials by air plasma treatment. \emph{Journal of Hospital Infection} \textbf{2010}, \emph{76}, 234--242\relax
\mciteBstWouldAddEndPuncttrue
\mciteSetBstMidEndSepPunct{\mcitedefaultmidpunct}
{\mcitedefaultendpunct}{\mcitedefaultseppunct}\relax
\EndOfBibitem
\bibitem[Zhao \latin{et~al.}(2012)Zhao, Lee, and Sen]{ZHAO201233}
Zhao,~L.~H.; Lee,~J.; Sen,~P.~N. Long-term retention of hydrophilic behavior of plasma treated polydimethylsiloxane (PDMS) surfaces stored under water and Luria-Bertani broth. \emph{Sensors and Actuators A: Physical} \textbf{2012}, \emph{181}, 33--42\relax
\mciteBstWouldAddEndPuncttrue
\mciteSetBstMidEndSepPunct{\mcitedefaultmidpunct}
{\mcitedefaultendpunct}{\mcitedefaultseppunct}\relax
\EndOfBibitem
\bibitem[Sarfati and Schwartz(2020)Sarfati, and Schwartz]{sarfati2020temporally}
Sarfati,~R.; Schwartz,~D.~K. Temporally anticorrelated subdiffusion in water nanofilms on silica suggests near-surface viscoelasticity. \emph{ACS nano} \textbf{2020}, \emph{14}, 3041--3047\relax
\mciteBstWouldAddEndPuncttrue
\mciteSetBstMidEndSepPunct{\mcitedefaultmidpunct}
{\mcitedefaultendpunct}{\mcitedefaultseppunct}\relax
\EndOfBibitem
\bibitem[Thomas \latin{et~al.}(2023)Thomas, Sinha~Mahapatra, Ganguly, and Tiwari]{Thomas202339}
Thomas,~T.~M.; Sinha~Mahapatra,~P.; Ganguly,~R.; Tiwari,~M.~K. Preferred Mode of Atmospheric Water Vapor Condensation on Nanoengineered Surfaces: Dropwise or Filmwise? \emph{Langmuir} \textbf{2023}, \emph{39}, 5396--5407, PMID: 37014297\relax
\mciteBstWouldAddEndPuncttrue
\mciteSetBstMidEndSepPunct{\mcitedefaultmidpunct}
{\mcitedefaultendpunct}{\mcitedefaultseppunct}\relax
\EndOfBibitem
\bibitem[Guo \latin{et~al.}(2012)Guo, Zhao, Bai, and Qiao]{GUO20129087}
Guo,~L.; Zhao,~X.; Bai,~Y.; Qiao,~L. Water adsorption behavior on metal surfaces and its influence on surface potential studied by in situ SPM. \emph{Applied Surface Science} \textbf{2012}, \emph{258}, 9087--9091\relax
\mciteBstWouldAddEndPuncttrue
\mciteSetBstMidEndSepPunct{\mcitedefaultmidpunct}
{\mcitedefaultendpunct}{\mcitedefaultseppunct}\relax
\EndOfBibitem
\bibitem[Yan \latin{et~al.}(1987)Yan, Meilink, Warren, and Wynblatt]{BenIEEE1987}
Yan,~B.-D.; Meilink,~S.; Warren,~G.; Wynblatt,~P. Water Adsorption and Surface Conductivity Measurements onalpha-Alumina Substrates. \emph{IEEE Transactions on Components, Hybrids, and Manufacturing Technology} \textbf{1987}, \emph{10}, 247--251\relax
\mciteBstWouldAddEndPuncttrue
\mciteSetBstMidEndSepPunct{\mcitedefaultmidpunct}
{\mcitedefaultendpunct}{\mcitedefaultseppunct}\relax
\EndOfBibitem
\bibitem[de~la Llave \latin{et~al.}(2012)de~la Llave, Molinero, and Scherlis]{Llave2012}
de~la Llave,~E.; Molinero,~V.; Scherlis,~D.~A. Role of Confinement and Surface Affinity on Filling Mechanisms and Sorption Hysteresis of Water in Nanopores. \emph{The Journal of Physical Chemistry C} \textbf{2012}, \emph{116}, 1833--1840\relax
\mciteBstWouldAddEndPuncttrue
\mciteSetBstMidEndSepPunct{\mcitedefaultmidpunct}
{\mcitedefaultendpunct}{\mcitedefaultseppunct}\relax
\EndOfBibitem
\bibitem[Gr{\"u}nberg \latin{et~al.}(2004)Gr{\"u}nberg, Emmler, Gedat, Shenderovich, Findenegg, Limbach, and Buntkowsky]{grunberg2004hydrogen}
Gr{\"u}nberg,~B.; Emmler,~T.; Gedat,~E.; Shenderovich,~I.; Findenegg,~G.; Limbach,~H.; Buntkowsky,~G. Hydrogen bonding of water confined in mesoporous silica MCM-41 and SBA-15 studied by $^{1}$H solid-state NMR. \emph{Chemistry} \textbf{2004}, \emph{10}, 5689--5696\relax
\mciteBstWouldAddEndPuncttrue
\mciteSetBstMidEndSepPunct{\mcitedefaultmidpunct}
{\mcitedefaultendpunct}{\mcitedefaultseppunct}\relax
\EndOfBibitem
\bibitem[Li \latin{et~al.}(2017)Li, Li, Wu, Feng, Zhang, and Zhang]{LI2017253}
Li,~J.; Li,~X.; Wu,~K.; Feng,~D.; Zhang,~T.; Zhang,~Y. Thickness and stability of water film confined inside nanoslits and nanocapillaries of shale and clay. \emph{International Journal of Coal Geology} \textbf{2017}, \emph{179}, 253--268\relax
\mciteBstWouldAddEndPuncttrue
\mciteSetBstMidEndSepPunct{\mcitedefaultmidpunct}
{\mcitedefaultendpunct}{\mcitedefaultseppunct}\relax
\EndOfBibitem
\bibitem[Freund \latin{et~al.}(1999)Freund, Halbritter, and Hörber]{Freund1999}
Freund,~J.; Halbritter,~J.; Hörber,~J. How dry are dried samples? Water adsorption measured by STM. \emph{Microscopy Research and Technique} \textbf{1999}, \emph{44}, 327--338\relax
\mciteBstWouldAddEndPuncttrue
\mciteSetBstMidEndSepPunct{\mcitedefaultmidpunct}
{\mcitedefaultendpunct}{\mcitedefaultseppunct}\relax
\EndOfBibitem
\bibitem[Brunauer \latin{et~al.}(1938)Brunauer, Emmett, and Teller]{BET1938}
Brunauer,~S.; Emmett,~P.~H.; Teller,~E. Adsorption of Gases in Multimolecular Layers. \emph{Journal of the American Chemical Society} \textbf{1938}, \emph{60}, 309--319\relax
\mciteBstWouldAddEndPuncttrue
\mciteSetBstMidEndSepPunct{\mcitedefaultmidpunct}
{\mcitedefaultendpunct}{\mcitedefaultseppunct}\relax
\EndOfBibitem
\bibitem[Naono and Hakuman(1991)Naono, and Hakuman]{NAONO1991405}
Naono,~H.; Hakuman,~M. Analysis of adsorption isotherms of water vapor for nonporous and porous adsorbents. \emph{Journal of Colloid and Interface Science} \textbf{1991}, \emph{145}, 405--412\relax
\mciteBstWouldAddEndPuncttrue
\mciteSetBstMidEndSepPunct{\mcitedefaultmidpunct}
{\mcitedefaultendpunct}{\mcitedefaultseppunct}\relax
\EndOfBibitem
\bibitem[Ouchi \latin{et~al.}(2014)Ouchi, Hamamoto, Mori, Takata, and Etoh]{OUCHI2014219}
Ouchi,~T.; Hamamoto,~Y.; Mori,~H.; Takata,~S.; Etoh,~A. Water vapor adsorption equilibrium and adsorption/desorption rate of porous alumina film adsorbent synthesized with anodization on heat transfer plate. \emph{Applied Thermal Engineering} \textbf{2014}, \emph{72}, 219--228, Special Issue: International Symposium on Innovative Materials for Processes in Energy Systems 2013 (IMPRES2013)\relax
\mciteBstWouldAddEndPuncttrue
\mciteSetBstMidEndSepPunct{\mcitedefaultmidpunct}
{\mcitedefaultendpunct}{\mcitedefaultseppunct}\relax
\EndOfBibitem
\bibitem[Butt \latin{et~al.}(2003)Butt, Graf, and Kappl]{book_interf}
Butt,~H.-J.; Graf,~K.; Kappl,~M. \emph{Physics \& Chemistry of Interfaces}; John Wiley and Sons, Ltd, 2003; Chapter 9, pp 177--205\relax
\mciteBstWouldAddEndPuncttrue
\mciteSetBstMidEndSepPunct{\mcitedefaultmidpunct}
{\mcitedefaultendpunct}{\mcitedefaultseppunct}\relax
\EndOfBibitem
\bibitem[Visser(1972)]{visser1972hamaker}
Visser,~J. On Hamaker constants: A comparison between Hamaker constants and Lifshitz-van der Waals constants. \emph{Advances in colloid and interface science} \textbf{1972}, \emph{3}, 331--363\relax
\mciteBstWouldAddEndPuncttrue
\mciteSetBstMidEndSepPunct{\mcitedefaultmidpunct}
{\mcitedefaultendpunct}{\mcitedefaultseppunct}\relax
\EndOfBibitem
\bibitem[Alam \latin{et~al.}(2014)Alam, Howlader, and Deen]{Alam_2014}
Alam,~A.~U.; Howlader,~M. M.~R.; Deen,~M.~J. The effects of oxygen plasma and humidity on surface roughness, water contact angle and hardness of silicon, silicon dioxide and glass. \emph{Journal of Micromechanics and Microengineering} \textbf{2014}, \emph{24}, 035010\relax
\mciteBstWouldAddEndPuncttrue
\mciteSetBstMidEndSepPunct{\mcitedefaultmidpunct}
{\mcitedefaultendpunct}{\mcitedefaultseppunct}\relax
\EndOfBibitem
\bibitem[Wang \latin{et~al.}(2021)Wang, Klein, Carnevale, and Borguet]{wang2021investigations}
Wang,~R.; Klein,~M.~L.; Carnevale,~V.; Borguet,~E. Investigations of water/oxide interfaces by molecular dynamics simulations. \emph{Wiley Interdisciplinary Reviews: Computational Molecular Science} \textbf{2021}, \emph{11}, e1537\relax
\mciteBstWouldAddEndPuncttrue
\mciteSetBstMidEndSepPunct{\mcitedefaultmidpunct}
{\mcitedefaultendpunct}{\mcitedefaultseppunct}\relax
\EndOfBibitem
\bibitem[Sarfati and Schwartz(2020)Sarfati, and Schwartz]{Sarfati-1021acsnano}
Sarfati,~R.; Schwartz,~D.~K. Temporally Anticorrelated Subdiffusion in Water Nanofilms on Silica Suggests Near-Surface Viscoelasticity. \emph{ACS Nano} \textbf{2020}, \emph{14}, 3041--3047, PMID: 31935060\relax
\mciteBstWouldAddEndPuncttrue
\mciteSetBstMidEndSepPunct{\mcitedefaultmidpunct}
{\mcitedefaultendpunct}{\mcitedefaultseppunct}\relax
\EndOfBibitem
\bibitem[Asay and Kim(2005)Asay, and Kim]{Asay2005}
Asay,~D.~B.; Kim,~S.~H. Evolution of the Adsorbed Water Layer Structure on Silicon Oxide at Room Temperature. \emph{The Journal of Physical Chemistry B} \textbf{2005}, \emph{109}, 16760--16763, PMID: 16853134\relax
\mciteBstWouldAddEndPuncttrue
\mciteSetBstMidEndSepPunct{\mcitedefaultmidpunct}
{\mcitedefaultendpunct}{\mcitedefaultseppunct}\relax
\EndOfBibitem
\bibitem[Gupta and Meuwly(2013)Gupta, and Meuwly]{C3FD00096F}
Gupta,~P.~K.; Meuwly,~M. Dynamics and vibrational spectroscopy of water at hydroxylated silica surfaces. \emph{Faraday Discuss.} \textbf{2013}, \emph{167}, 329--346\relax
\mciteBstWouldAddEndPuncttrue
\mciteSetBstMidEndSepPunct{\mcitedefaultmidpunct}
{\mcitedefaultendpunct}{\mcitedefaultseppunct}\relax
\EndOfBibitem
\bibitem[Verdaguer \latin{et~al.}(2007)Verdaguer, Weis, Oncins, Ketteler, Bluhm, and Salmeron]{doi:10.1021/la700893w}
Verdaguer,~A.; Weis,~C.; Oncins,~G.; Ketteler,~G.; Bluhm,~H.; Salmeron,~M. Growth and Structure of Water on SiO2 Films on Si Investigated by Kelvin Probe Microscopy and in Situ X-ray Spectroscopies. \emph{Langmuir} \textbf{2007}, \emph{23}, 9699--9703, PMID: 17696552\relax
\mciteBstWouldAddEndPuncttrue
\mciteSetBstMidEndSepPunct{\mcitedefaultmidpunct}
{\mcitedefaultendpunct}{\mcitedefaultseppunct}\relax
\EndOfBibitem
\bibitem[Argyris \latin{et~al.}(2011)Argyris, Ho, Cole, and Striolo]{ArgyrisJCP2011}
Argyris,~D.; Ho,~T.; Cole,~D.~R.; Striolo,~A. Molecular Dynamics Studies of Interfacial Water at the Alumina Surface. \emph{The Journal of Physical Chemistry C} \textbf{2011}, \emph{115}, 2038--2046\relax
\mciteBstWouldAddEndPuncttrue
\mciteSetBstMidEndSepPunct{\mcitedefaultmidpunct}
{\mcitedefaultendpunct}{\mcitedefaultseppunct}\relax
\EndOfBibitem
\end{mcitethebibliography}

\end{document}